\documentclass{elsart}
\usepackage{amssymb}
\usepackage{amsmath}
\usepackage{graphicx}
\usepackage{longtable}
\input{Qcircuit}

\renewcommand{\ket}[1]{\ensuremath{|#1\rangle}}
\renewcommand{\bra}[1]{\ensuremath{\langle#1|}}
\newcommand{\inner}[2]{\ensuremath{\langle#1|#2\rangle}}

\newcommand{\NOT}{\textsc{not}}
\newcommand{\XOR}{\textsc{xor}}
\newcommand{\NAND}{\textsc{nand}}
\newcommand{\CNOT}{controlled-\NOT}
\newcommand{\CCNOT}{controlled-\CNOT}
\newcommand{\TOFFOLI}{\textsc{toffoli}}

\newcommand{\SWAP}{\textsc{swap}}
\newcommand{\FREDKIN}{\textsc{fredkin}}
\newcommand{\GRAPE}{\textsc{grape}}
\newcommand{\third}{\mbox{$\frac{1}{3}$}}
\newcommand{\smhalf}{\raisebox{0.4ex}{$\scriptstyle\frac{1}{2}$}}
\newcommand{\eg}{\textit{e.g.}}

\begin{document}

\begin{frontmatter}

\title{Quantum Computing with NMR}
\author{Jonathan A. Jones}
\address{Centre for Quantum Computation, Clarendon Laboratory,
University of Oxford, Parks Road, Oxford OX1~3PU, UK}
\address{Centre for Advanced ESR,
University of Oxford, South Parks Road, Oxford OX1~3QR, UK}
\ead{jonathan.jones@qubit.org}
\received\today
\tableofcontents

%\begin{abstract}
% Text of abstract
%\end{abstract}

%\begin{keyword}
%\end{keyword}
\end{frontmatter}
%\tableofcontents

% main text
\section{Introduction}

Quantum computers \cite{Deutsch1985,Bennett2000,Knill2010,Ladd2010} are explicitly quantum mechanical systems that use phenomena such as superposition and entanglement to perform computational tasks more efficiently than any classical computer \cite{Feynman1982}. More generally, quantum information technologies can perform information processing tasks, such as unconditionally secure communication, which cannot be achieved by classical systems. Unsurprisingly quantum computation has generated an enormous amount of interest, reflecting not just its potential technological importance, but also the intellectual importance of the challenge it provides to previous formulations of computational complexity theory \cite{ARbook}, built on the Church--Turing thesis, which asserts that all physical models of computation are essentially equivalent to one another in computational power.

This intrinsic interest is partly tempered by the great practical difficulty in building large scale devices capable of implementing technologically important computations, but it has at least proved fairly simple to build small demonstration devices, and NMR has played a leading role in this field for many years.  The first ideas on how to build quantum computers with NMR \cite{Cory1996,Cory1997,Cory1998a,Gershenfeld1997,Chuang1998b} were swiftly followed by the first implementations of quantum algorithms \cite{Jones1998c,Chuang1998}, and it is only recently that other techniques have achieved similar sophistication in the manipulation of quantum states: although an implementation of the fundamental \CNOT\ gate with trapped ions was demonstrated in 1995 \cite{Monroe1995} it was not until 2003 that a complete implementation of Deutsch's algorithm was achieved \cite{Gulde2003}.  It must, however, be remembered that the great difficulty in preparing NMR systems in pure spin states, reflecting the tiny energy gap between nuclear spin states, has given rise to grave concerns about the direct relevance of NMR techniques to attempts to build large scale devices \cite{Warren1997,Gershenfeld1997a}, and has even led to questioning of whether many NMR quantum computations can be considered true quantum computations at all \cite{Braunstein1999}.

\subsection{Structure and scope}
In my previous introduction \cite{Jones1998b} and reviews \cite{Jones2000a,Jones2001a,Jones2001} I described the fundamental concepts of quantum computation, and how these are related to more familiar concepts in conventional NMR spectroscopy.  For the most part I will only briefly repeat these here, and will largely assume familiarity with these basic concepts.  A large number of excellent textbooks, \eg\ \cite{NCbook,Merminbook,Morschbook,SSbook2e}, have now been published, providing many different approaches to the necessary background with almost any desired style and level of sophistication, and a number of reviews, \eg\ \cite{Mehring1999,Cory2000,Vandersypen2004,Heidebrecht2006,Ryan2008,Suter2008}, have described both fundamental concepts and recent developments in particular areas of NMR quantum computing.

%\subsection{The DiVincenzo criteria}
The general requirements for implementing quantum information processing on a physical system are usually considered in terms of the seven DiVincenzo criteria \cite{Divincenzo2000,Divincenzo2001}, where the first five criteria are required for quantum computation and the final two permit quantum computers to be connected by quantum information links, in effect building a quantum internet.  The first five criteria are:
\begin{enumerate}
\item a scalable physical system with well characterized qubits;
\item the ability to initialize the state of the qubits to a simple fiducial state such as \ket{000\dots};
\item long relevant decoherence times, much longer than the gate operation time;
\item a ``universal'' set of quantum gate;
\item a qubit-specific measurement capability.
\end{enumerate}
As is traditional I will structure the first part of my discussion around these five criteria, but will take them in a different, and more helpful, order.  I will not consider the two additional criteria, the interconversion of \textit{static qubits} (suitable for information processing) and \textit{flying qubits} (suitable for information transmission), and the transmission of these flying qubits from place to place.

I will only describe in detail the simplest method for implementing quantum information processing with NMR, that is using spin-$\half$ nuclei in molecules in solution, although I will briefly comment on alternative approaches including solid-state NMR, the use of liquid crystal solvents, and the use of quadrupolar nuclei, as well as on approaches involving electron spins, either alone or combined with nuclear spins.  As the discussion is structured around the DiVincenzo criteria, these topics will inevitably be scattered around the text.  I will not discuss exotic long range proposals, such as the famous Kane proposal \cite{Kane1998} for a large scale quantum computer based on ENDOR manipulation of \nuc{31}{P} spins in doped silicon, but will largely confine myself to methods which can be implemented with relatively conventional NMR and ESR spectrometers, and for which at least elementary experimental demonstrations of quantum information processing have in fact been performed.  These sections are followed by an introduction to some quantum algorithms and communication protocols, highlighting some experimental achievements.  The available literature is now too large to include every paper, or even every topic that has been investigated, and my selection is inevitably personal and incomplete.

\section{Qubits}
A quantum bit, or \textit{qubit} \cite{Schumacher1995}, is simply a two level quantum system, where the two levels are labelled as \ket{0} and \ket{1}, effectively identifying them with the two logic states $0$ and $1$.  This choice is usually called the computational basis, and is the implicit basis used here; the state \ket{0} is usually chosen to have lower energy than \ket{1} but this is not essential.

\subsection{Superpositions and entanglement}
As a qubit is a quantum system it is not confined to these two basis states, but can exist in a general superposition state of the form
\begin{equation}\label{eq:qubit}
\alpha\ket{0}+\beta\ket{1}
\end{equation}
where $\alpha$ and $\beta$ are arbitrary complex numbers subject to the constraint that $|\alpha|^2+|\beta|^2=1$.  However, this traditional description of a superposition state of a qubit is in some sense excessive, in that there is an ambiguity in the \textit{global phase} of the state \cite{Diracbook}, so that the set of states
\begin{equation}\label{eq:qubitgp}
\textrm{e}^{\textrm{i}\gamma}\ket{\psi}=\alpha\textrm{e}^{\textrm{i}\gamma}\ket{0}+\beta\textrm{e}^{\textrm{i}\gamma}\ket{1}
\end{equation}
(where $\gamma$ is the global phase, taking real values) are completely indistinguishable from one another.  Equivalently, the global phase can be incorporated into the value of one of the coefficients, usually taken as the first coefficient $\alpha$, so that this coefficient can be assumed to be real.  Such global phases must be sharply distinguished from the \textit{relative phase} of two coefficients in a superposition, which is a measurable quantity.

The ability of quantum computers to process data in superposition states lies at the heart of their power.  One particularly important pair of superposition states is \ket{+} and \ket{-}, defined by
\begin{equation}\label{eq:ketpm}
\ket{\pm}=\frac{\ket{0}\pm\ket{1}}{\sqrt{2}},
\end{equation}
which play a key role in many quantum algorithms.

A single qubit is not a particularly interesting object, but even a two qubit system can show interesting computational behaviour.  A pair of qubits has four basis states, and can be found in some general superposition of all four,
\begin{equation}\label{eq:qubits}
\alpha\ket{00}+\beta\ket{01}+\gamma\ket{10}+\delta\ket{11}
\end{equation}
with the constraint $|\alpha|^2+|\beta|^2+|\gamma|^2+|\delta|^2=1$.  These general states can be divide into two major groups: \textit{product states}, which can be written as the direct product of two terms of the form Eq.~\ref{eq:qubit}, such as
\begin{equation}
\frac{\ket{00}+\ket{01}}{\sqrt{2}}=\ket{0}\ket{+},%\otimes\frac{\ket{0}+\ket{1}}{\sqrt{2}},
\end{equation}
and \textit{entangled states}, which cannot be decomposed in this way, such as the states
\begin{equation}\label{eq:Bell}
\ket{\phi_\pm}=\frac{\ket{00}\pm\ket{11}}{\sqrt{2}}\quad\ket{\psi_\pm}=\frac{\ket{01}\pm\ket{10}}{\sqrt{2}}
\end{equation}
which are the four maximally entangled Bell-states.  Entanglement is an important, but surprisingly slippery, concept which will be returned to on a number of occasions.  The slipperiness arises from the fact that although it is relatively easy to demonstrate that a state is a product state, by writing it down in explicit product form, it is much more difficult to prove that a state cannot possibly be written in a product form, and so apparently entangled states may in fact turn out to be product states.

\subsection{Pure states and mixed states}
A qubit in a well defined quantum state \ket{\psi} (which can be a computational basis state or a superposition state) is most conveniently described in the ket notation used above, but such a state, called a \textit{pure state}, can also be described using the density matrix description
\begin{equation}\label{eq:qubitrho}
\rho=\ket{\psi}\bra{\psi}=\begin{pmatrix}\alpha\\\beta\end{pmatrix}\times\begin{pmatrix}\alpha^*&\beta^*\end{pmatrix}=\begin{pmatrix}\alpha\alpha^*&\alpha\beta^*\\\beta\alpha^*&\beta\beta^*\end{pmatrix}
\end{equation}
where the constraint on $\alpha$ and $\beta$ now appears as the property that
\begin{equation}\label{eq:trrho}
\textrm{tr}(\rho)=|\alpha|^2+|\beta|^2=1.
\end{equation}
For a pure state the density matrix $\rho$ has the property that
\begin{equation}\label{eq:rho2}
\rho^2=\ket{\psi}\inner{\psi}{\psi}\bra{\psi}=\ket{\psi}\bra{\psi}=\rho
\end{equation}
so that $\rho$ is said to be idempotent, and $\textrm{tr}(\rho^2)=1$.  A qubit can also be in a more general \textit{mixed state}, which is a statistical mixture of pure states
\begin{equation}\label{eq:rhomix}
\sum_ip_i\ket{\psi_i}\bra{\psi_i}
\end{equation}
where the $p_i\ge0$ are probabilities, so that $\sum_ip_i=1$.  Density matrices for mixed states are not idempotent.  Mixed states can be classified according to their \textit{purity}, which can be defined as $\textrm{tr}(\rho^2)\le1$, although other definitions are in use.  Unitary transformations of a state leave the purity unchanged, while non-unitary operations normally act to to reduce the purity; increasing the purity of a state is a particularly tricky operation.

The ambiguous global phase term is not present in the density matrix description of a quantum state: for the state considered above, Eq.~\ref{eq:qubitgp}, the corresponding density matrix is
\begin{equation}\label{eq:rhogp}
\textrm{e}^{\textrm{i}\gamma}\ket{\psi}\bra{\psi}\textrm{e}^{-\textrm{i}\gamma}=\ket{\psi}\bra{\psi}
\end{equation}
as the contributions from the global phase terms in the ket and bra cancel out.  Thus the density matrix contains only the physically observable information about a state.

%Density matrices are a useful description of a quantum state as they encode all the physically accessible properties of the state.
For a pure state, the ket and density matrix approaches give two different ways of relating qubit states to the Bloch sphere.  Firstly, considering the two constraints on the state of a qubit, Eq.~\ref{eq:qubit}, it can be written as
\begin{equation}\label{eq:qubitbs}
\cos(\theta/2)\ket{0}+\sin(\theta/2)\textrm{e}^{\textrm{i}\phi}\ket{1}
\end{equation}
where $0\le\theta\le\pi$, which determined the relative amplitude of the two basis states, is a co-latitude, and $0\le\phi<2\pi$, the relative phase, is an azimuth angle.  These two angles define a unique point on a sphere of radius 1, the Bloch sphere.  Alternatively, the density matrix description of a qubit, Eq.~\ref{eq:qubitrho} can be written in the Pauli basis (which extends the three traditional Pauli matrices into a matrix basis by including $\sigma_0$, the two by two identity matrix) as
\begin{equation}\label{eq:qubitrhobs}
\half\left(s_0\sigma_0+s_x\sigma_x+s_y\sigma_y+s_z\sigma_z\right).
\end{equation}
As both the density matrix and the Pauli basis are manifestly Hermitian, the four coefficients $s_0$, $s_x$, $s_y$ and $s_z$ are necessarily real.  The constraint $|\alpha|^2+|\beta|^2=1$ sets $s_0=1$, leaving the remaining three coefficients which define a vector of length 1. A similar approach can be used for mixed states, but in this case the state is described by a point inside the Bloch sphere.  This is most easily seen by noting that the Bloch vector for a mixed state will be a weighted vector sum of the contributing Bloch vectors, and so must have length less than one.

In the discussions above I have used an explicit description of a density matrix, Eq.~\ref{eq:rhomix}, composing it as a particular mixture of pure states.  It is tempting to conclude that a given decomposition has some particular meaning, but in fact mixed state density matrices do not have unique decompositions, and no decomposition is more correct than any other.  One famous example is that an equally weighted mixture of \ket{+} and \ket{-} states is indistinguishable from an equally weighted mixture of \ket{0} and \ket{1} states:
\begin{equation}\label{eq:pmmix}\begin{split}
\half\ket{+}\bra{+}+\half\ket{-}\bra{-}&=\half\begin{pmatrix}\half&\half\\\half&\half\end{pmatrix}+\half\begin{pmatrix}\half&-\half\\-\half&\half\end{pmatrix}\\
&=\begin{pmatrix}\half&0\\0&\half\end{pmatrix}=\half\ket{0}\bra{0}+\half\ket{1}\bra{1}.
\end{split}\end{equation}
As the density matrix description contains all physically accessible properties of a state, these two mixtures are essentially equivalent.

The belief that one particular decomposition of a mixed state has particular validity is sometimes called the \textit{preferred ensemble fallacy} or \textit{partition ensemble fallacy} \cite{Kok2000}, and can lead to serious confusion.  This does not, of course, mean that it is not permissible to use a particular decomposition; indeed, it is frequently very useful to do so. However, one must be extremely cautious when making arguments that depend explicitly on any particular decomposition.  For example, it is tempting to argue that a mixture of entangled states must itself be entangled, but it is easily shown that the apparently entangled mixture
\begin{equation}
\half\ket{\phi_+}\bra{\phi_+}+\half\ket{\phi_-}\bra{\phi_-}
\end{equation}
can also be decomposed as the mixture
\begin{equation}
\half\ket{00}\bra{00}+\half\ket{11}\bra{11},
\end{equation}
which is clearly a mixture of product states.  Such states are not entangled in any useful sense; this point will be explored in more detail when I consider the initialisation of quantum states.

\subsection{Spin-$\half$ nuclei in liquid samples}
There are many possible physical implementations of a qubit, frequently based on a two-level subsystem of a larger quantum system, but a particularly natural implementation is provided by a spin-$\half$ atomic nucleus, such as \nuc{1}{H}, and this is the approach used in the great majority of NMR implementations. A two-qubit quantum computer can be built from two spin-$\half$ nuclei, and so on.  As will be discussed below it is necessary that the two nuclei are distinct, so that the two qubits can be separately addressed, and there must be some sort of spin--spin interaction, so that two-qubit logic gates can be constructed.  This is easily achieved by using two inequivalent nuclei in a molecule.  The great majority of experiments have been performed using conventional liquid state samples, where rapid molecular tumbling greatly simplifies the spin Hamiltonian \cite{EBWbook}, with the isotropic part of the scalar J-coupling providing a suitable form of coupling.  Such a sample contains many copies of each molecule, but as intermolecular interactions are cancelled out by tumbling the result is an ensemble of identical independent molecules, which can for most practical purposes be treated as a single molecule in a mixed spin state.

In systems with two or more qubits an important practical distinction can be made between systems where all the spins are different nuclear species (a fully \textit{heteronuclear} spin system) and those with two or more nuclei of the same type (a partially  \textit{homonuclear} spin system).  As there are a limited number of different spin-$\half$ nuclei, among which only six (\nuc{1}{H}, \nuc{13}{C}, \nuc{15}{N}, \nuc{19}{F}, \nuc{29}{Si} and \nuc{31}{P}) have been used in quantum computing experiments it is clear than only very small quantum computers can be fully heteronuclear, but the relative ease of working with such systems makes them popular for implementing simple tasks.  Other spin-$\half$ nuclei, such as \nuc{117}{Sn} and \nuc{119}{Sn} could in principle be used, but have not yet found use, presumably reflecting their low gyromagnetic ratios, low natural abundances, and the relative difficulty in obtaining isotopically labelled compounds.  Similarly the difficulty of obtaining NMR probes capable of addressing more than five nuclei simultaneously (including the \nuc{2}{H} lock), and the ready availability of sensitive probes optimised for particular nuclear combinations (such as \nuc{1}{H}, \nuc{13}{C} and \nuc{15}{N}, used in NMR studies of proteins) has to some extent determined the choice of spin systems used so far.

A (very incomplete) list of systems of spin-$\half$ nuclei which have been used to implement NMR quantum computing is given in table~\ref{tab:spins}.  Spins used to study single isolated qubits are not listed; nor are systems involving high-spin nuclei.  A variety of larger spin systems have been explored in histidine \cite{Negrevergne2006} which contains 14 spin-$\half$ nuclei.

\newpage
%\begin{table}
%\begin{tabular}{|l|l|l|l|}
\begin{longtable}{|l|l|l|l|}
\caption{A partial list of spin systems of spin-$\half$ nuclei used to implement quantum computing with NMR; only very selective references are given for each spin system. Isotopic labelling is not indicated, and in some cases different sized spin systems are created either by selective isotopic labelling of a compound, or by using decoupling to effectively remove a spin from the spin system.}\label{tab:spins}\\
\hline
$n$&spins&molecule&references\\\hline
2&HH&2,3-dibromothiophene&\cite{Cory1996,Cory1998a,Gershenfeld1997,Somaroo1999,Tseng2000}\\
 &&cytosine&\cite{Jones1998c,Jones1998d,Jones1999a}\\
 &&uracil&\cite{Linden1999b}\\
 &&5-nitrofuraldehyde&\cite{Dorai2000a,Arvind2001,Mahesh2001}\\
 &&coumarin&\cite{Dorai2000a}\\
 &&Ir(H)Cl(H)(CO)(PPh$_3$)$_2$&\cite{Hubler2000}\\
 &&Ru(H)$_2$(CO)$_2$(dppe)&\cite{Anwar2004,Anwar2004b}\\
 &&Ru(H)$_2$(CO)$_2$(dpae)&\cite{Anwar2004a}\\
 &&5-bromothiophene-2-carbaldehyde&\cite{Roy2010}\\
 \hline
 &HC&chloroform&\cite{Chuang1998,Chuang1998a,Knill1998,Yannoni1999,Marjanska2000,Jones2000b,Childs2001}\\
 &&formate&\cite{Leung1999,Xiao2005,Xiao2006a}\\
 &&benzene&\cite{Zhang1999}\\
 &&dimethylformamide&\cite{Das2003a}\\
 \hline
 &HF&5-fluorouracil&\cite{Dorai2001}\\
 \hline
 &HP&phosphonic acid&\cite{Fu1999,Long2001a}\\
\hline
 &PP&1,4-diphosphafulvene&\cite{Ito2009}\\
 \hline
3&HHH&2,3-dibromopropionic acid&\cite{Linden1998,Linden1999a}\\
 &&chlorostyrene&\cite{Du2000a,Du2001a}\\
 \hline
 &HHF&4-fluoro-7-nitro-benzofuran&\cite{Dorai2000a,Das2004}\\
 \hline
 &HHP&\textit{E}-(2-chloroethenyl)phosphonic acid&\cite{Cummins2002}\\
 \hline
 &HCC&trichlororethene&\cite{Nielsen1998,Laflamme1998,Cory1998}\\
 &&tris(trimethylsilyl)silane-acetylene&\cite{Henry2006}\\
 \hline
 &HCF&dibromofluoromethane&\cite{Vandersypen2000}\\
 &&diethyl-fluoromalonate&\cite{Peng2008}\\
 \hline
 &CCC&alanine&\cite{Cory1998,Nelson2000,Collins2000,Kim2000,Viola2001,Weinstein2001,Teklemariam2001,Teklemariam2002,Teklemariam2003,Fortunato2002,Weinstein2002,Kim2002,Xiao2002,Lee2002a,Fitzsimons2007}\\
 \hline
 &FFF&bromotrifluoroethene&\cite{Vandersypen1999}\\
 &&2,3,4-trifluoroaniline&\cite{Mangold2004}\\
 &&iodotrifluoroethene&\cite{Du2007,Mitra2008}\\
 \hline
  &PPP&3,4-dihydro-1,3,4-triphosphacyclopenta[\textit{a}]indene&\cite{Ito2009}\\
\hline
4&HHHH&1-chloro-2-nitrobenzene&\cite{Cory1998a}\\
\hline
 &HHFF&2,3-difluoro-6-nitrophenol&\cite{Mahesh2001,Das2004,Das2004a}\\
 \hline
 &HCCC&alanine&\cite{Du2001a}\\
 \hline
 &HNCC&glycine&\cite{Ollerenshaw2003}\\
 \hline
 &CCCC&crotonic acid&\cite{Boulant2002,Boulant2003}\\
 \hline
 &FFFF&3-chloro-2,4,5,6-tetrafluoropyridine&\cite{Fung2001}\\
\hline
5&HHFFF&2,4,5-trifluorobenzonitrile&\cite{Fung2001,Ermakov2003}\\
 \hline
 &HNCCF&glycine fluoride&\cite{Marx2000}\\
 \hline
 &HFFFF&2,3,4,6-tetrafluoroaniline&\cite{Kawamura2005,Kawamura2006,Kawamura2007}\\
 \hline
 &FFFFF&pentafluorobutadienyl complex&\cite{Vandersypen2000a}\\
 \hline
6&HHHHHH&inosine&\cite{Linden1999}\\
 \hline
7&HHHCCCC&crotonic acid&\cite{Knill2000,Knill2001}\\
 \hline
 &CCFFFFF&pentafluorobutadienyl complex&\cite{Vandersypen2001}\\
 \hline
%12&HHHHCCCCCCNN&histidine&\cite{Negrevergne2006}\\
%\hline
%\end{tabular}
%\end{table}
\end{longtable}
\newpage

One complexity which arises in larger spin systems is that they frequently include groups of two or more equivalent nuclei, such as the three \nuc{1}{H} nuclei in a methyl group, and these cannot be treated as independent qubits, but must all be considered together.  However, it is possible to decompose a methyl group as the combination of a fictitious spin-$\frac{3}{2}$ particle and a fictitious spin-$\half$ particle, and in some experiments it is possible to use only the spin-$\half$ component of a $\textrm{CH}_3$ group \cite{Knill2000,Knill2001} or the spin-$1$ component of a $\textrm{CH}_2$ group  \cite{Negrevergne2006}.  Equivalent nuclei can also be used directly in situations where only partial control of the spin system is needed, such as entanglement assisted magnetic field sensing \cite{Jones2009,Simmons2010}.

\subsection{Product operators and deviation density matrices}
Liquid state NMR experiments are usually described using product operator notation \cite{Sorensen1983}, which is almost equivalent to the Pauli basis, differing only in normalisation.  As NMR spin states are almost always highly mixed it is common practice to be quite casual about normalisation, and simply concentrate on the relative size of terms in the density matrix which can be measured.  This approach can lead to serious errors when describing quantum computing, and it is necessary to proceed with care.

We can describe single qubit density matrices using product operators; for example the pure ground state of a single qubit is described by
\begin{equation}
\ket{0}\bra{0}=\begin{pmatrix}1&0\\0&0\end{pmatrix}=\begin{pmatrix}\half&0\\0&\half\end{pmatrix}+\begin{pmatrix}\half&0\\0&-\half\end{pmatrix}=\half{E}+I_z
\end{equation}
while $\ket{1}\bra{1}=\half{E}-I_z$.  As NMR observables are traceless the $\half{E}$ component is not directly detectable, and so we can say that $\ket{0}\bra{0}$ is equivalent to $I_z$.  This description, in which multiples of the density matrix are ignored, is sometimes called the \textit{deviation density matrix} \cite{Chuang1998b} description of a state.  Note that deviation density matrices are not proper density matrices, in that they have trace 0 and negative eigenvalues, and so cannot actually describe the state of a real physical system; they are simply a useful computational fiction.  Furthermore, the thermal state of a single spin is in fact the mixed state $\half{E}+\delta I_z$, where $\delta$ is the polarisation of the spin.  Although the absolute value of $\delta$ is not normally directly measurable, it will nevertheless ultimately determine the signal size, and so cannot be entirely ignored.

For a single spin the thermal state is broadly equivalent to the desired initial state $\ket{0}\bra{0}$, but this simple equivalence does not remain true for larger spin systems.  For example, with two spins the desired initial state is
\begin{equation}\label{eq:pptwopo}
\ket{00}\bra{00}=\begin{pmatrix}1&0&0&0\\0&0&0&0\\0&0&0&0\\0&0&0&0\end{pmatrix}=\half(\half{E}+I_z+S_z+2I_zS_z)
\end{equation}
while the thermal state for a homonuclear spin system is $\half{E}+\delta(I_z+S_z)$.  These points are explored in detail in Section~\ref{sec:initialisation} where methods of initialising spin systems are considered.

\subsection{Dipolar couplings}
NMR studies in the solid state \cite{Slichter1990,Schmidt-Rohr1994a} are considerably more complicated than those in the liquid state as the direct dipolar coupling term in the spin Hamiltonian is not averaged by rapid molecular tumbling.  This has the potential advantage that direct dipolar couplings are normally considerably larger than the isotropic part of the scalar coupling, but this on turn means that dipolar couplings are unlikely to be in the weak coupling limit, complicating the Hamiltonian further.  More seriously, intermolecular dipolar couplings mean that it is not usually possible to treat the molecules as an ensemble of identical independent copies.  Further difficulties arise from the orientation dependence of chemical shifts and the presence of multiple molecules with different orientations in the unit cells of many crystals.  Nevertheless, the considerable advantages that solid state systems might offer \cite{Cory2000}, most notably the ability to cool nuclear spins close to their ground states, have encouraged initial attempts at this problem.

The first implementation of quantum, computing by solid state NMR \cite{Leskowitz2003} used a single crystal of glycine, containing 10\% of fully \nuc{13}{C} and \nuc{15}{N} labelled glycine diluted with natural abundance glycine to minimise the effects of intermolecular couplings.  The sample orientation was set so that one molecule in the unit cell gave sharp lines well resolved from the signals from other orientations, and \nuc{1}{H} decoupling was applied throughout all experiments, effectively removing the \nuc{1}{H} nuclei from the spin system with the exception of an initial cross-polarization phase to increase the polarization of the \nuc{13}{C} and \nuc{15}{N} nuclei.  The heteronuclear couplings between \nuc{13}{C} and \nuc{15}{N} are weak compared to the Zeeman energy scale, and the homonuclear coupling between the two \nuc{13}{C} nuclei (conventionally called Co and C$\alpha$) can also be treated (to a reasonable approximation) as weak.  Thus the final spin system is very similar to that used in liquid state experiments, and similar techniques can be applied \cite{Leskowitz2003}.

This early work overcame the complexities inherent in solid state NMR by carefully choosing a spin system where they could be largely ignored rather than by tackling them head on.  A later more ambitious approach \cite{Baugh2005,Baugh2006,Ryan2008a} used three \nuc{13}{C} nuclei in fully \nuc{13}{C} labelled malonic acid.  Once again an oriented single crystal was used  containing 3.2\% labelled malonic acid diluted with natural abundance malonic acid.  In this spin system, however, the dipolar couplings cannot be taken as weak, and it is necessary to consider the full dipolar Hamiltonian in designing logic gates (the couplings are weak enough that it is still possible to identify the nuclear eigenstates with the computational basis states in a reasonably straightforward manner).  This problem was tackled by using strongly modulated composite pulses \cite{Fortunato2002} and \GRAPE\ (gradient ascent pulse engineering) pulses \cite{Khaneja2005} which will be described in Sections \ref{sec:SMCP} and \ref{sec:GRAPE}.

The methods described above all use oriented single crystals, as couplings which are truncated by the Zeeman interaction depend strongly on the relative orientation of the coupling tensor (aligned with the crystal axes) and the magnetic field.  The approach commonly used in conventional solid state NMR of spinning out these interactions and then selectively reintroducing them by recoupling \cite{Schmidt-Rohr1994a} will not work well in NMR quantum computing as the orientation dependence of the reintroduced couplings, visible as spinning sidebands, will prevent the accurate implementation of quantum gates.  A radically different approach, representing quantum information using Floquet states in spinning samples has been proposed \cite{Ding2001}, but has not been further explored.

NMR studies of molecules dissolved in liquid crystal solvents \cite{Emsley1975,Schmidt-Rohr1994} provide a convenient half-way house between conventional solutions and solid state samples: the partial local ordering induced by the solvent results in intramolecular dipolar couplings being scaled down, but not averaged to zero, while intermolecular couplings are effectively averaged, as in true liquids, and can be ignored.  Thus the system can again be treated as an ensemble of identical independent molecules.  Early attempts to implement NMR quantum computation with liquid crystals used systems where the internal couplings could still be treated as weak \cite{Yannoni1999,Marjanska2000,Fung2001}, and the experiments are very similar to conventional liquid state experiments, but later experiments have used systems with strongly coupled spins \cite{Das2004b,Mahesh2006}.

A more extreme approach is to use clusters of equivalent spins; the dipolar coupling between such spins cannot normally be seen in liquids, but is revealed in liquid crystal solutions \cite{Emsley1992}.  Early experiments were performed on the three \nuc{1}{H} nuclei and single \nuc{13}{C} nucleus in a methyl group \cite{Fung2001a,Mahesh2002}, but these have now been extended to much larger spin systems, most notably the 12 spins (six equivalent \nuc{1}{H} nuclei and six equivalent \nuc{13}{C} nuclei) in \nuc{13}{C} labelled benzene \cite{Lee2004,Lee2005,Lee2005a}.  A partial review of experiments in this field has now been published \cite{Lee2007}, and there has also been significant theoretical work \cite{Furman2006}.

\subsection{Quadrupolar nuclei}
Nuclear spins with spin quantum numbers greater than one half have quadrupolar interactions with electric field gradients \cite{Pound1950,Das1958} in addition to the Zeeman interaction with magnetic fields and dipolar and scalar couplings, and so are called \textit{quadrupolar nuclei}.  In the liquid state these terms are averaged by molecular motion, although their effects are frequently seen in the short relaxation times of many quadrupolar nuclei.  In solid state samples quadrupolar interactions can be comparable to or even larger than Zeeman interactions and have huge effects on the NMR spectra.  As before, the use of liquid crystal solvents provides a convenient middle way, allowing the quadrupolar interaction to be introduced in a measured and controllable fashion, although short relaxation times remain a problem.

More fundamentally still, such nuclei have more than two spin states and so do not provide a natural implementation of a qubit (although the behaviour of direct NMR observables can often be described using a Bloch sphere picture, more detailed consideration shows that the underlying behaviour is considerably more complicated).  Two approaches are then possible: either the spin can be used to represent a larger quantum information carrier (for example, a spin-$1$ nucleus such as \nuc{2}{H} is a quantum three-level system and so can be used to implement a qutrit \cite{Das2003c}), or a group of levels can be used to represent two or more quantum bits.  This second approach has dominated, with the four levels of the spin-$\frac{3}{2}$ nuclei \nuc{23}{Na} \cite{Khitrin2000,Ermakov2002,Sarthour2003,Das2003b,Bonk2004,Bonk2005,Bulnes2005,Teles2007} and \nuc{7}{Li} \cite{Sinha2001} being used to represent two qubits, and the eight levels in the spin-$\frac{7}{2}$ nucleus \nuc{113}{Cs} being used to represent three qubits \cite{Khitrin2001,Murali2002,Das2006,Gopinath2008}.  These experiments have used liquid crystal solvents, but studies have also been performed on \nuc{23}{Na} nuclei in a single crystal of $\textrm{NaNO}_3$ \cite{Kampermann2005}.

A major disadvantage of approaches of this kind is that embedding multiple qubits in a single spin can make it very hard to scale up the system.  In conventional quantum computing another qubit can be added to the system simply by adding another spin-$\half$ particle which is coupled to the existing spin system.  By contrast, to move from two qubits to three requires replacing a spin-$\frac{3}{2}$ nucleus with a spin-$\frac{7}{2}$ nucleus, while a four-qubit system would require a (non-existent) spin-$\frac{15}{2}$ nucleus.  This weakness can to some extent be overcome by using molecular magnets, which use clusters of spins to create systems equivalent to very high spin nuclei \cite{Leuenberger2001,Ardavan2007}, but this does not really address the fundamental problem.

\subsection{Electron spins: ESR and ENDOR}
There have been many proposals for implementing quantum information processing with electrons, or electrons and nuclei, in solid state systems \cite{Burkard2000}.  Here I will only consider a small number of systems which have been studied with relatively conventional ESR techniques \cite{Schweiger2001}.

One system which has been extensively studied \cite{Mehring2004,Morton2005,Morton2005a,Morton2006,Morton2006a,Ardavan2007a,Scherer2008} is the $^4\textrm{S}_{3/2}$ electronic state of nitrogen endrohedrally doped into a $\textrm{C}_{60}$ fullerene cage (\textit{i}-$\textrm{NC}_{60}$, or N@$\textrm{C}_{60}$), which provides a spin-$\frac{3}{2}$ system coupled to either a spin-$\half$ \nuc{15}{N} nucleus or a spin-$1$ \nuc{14}{N} nucleus.  However, although several quantum phenomena have been demonstrated this system has not yet been used for implementing full quantum computation.

An alternative spin system, which should perhaps be better suited to simple quantum computations, is provided by the single electron ($S=\half$) and \nuc{1}{H} nuclear spin ($I=\half$) in the $^{\bullet}\textrm{CH}$ radical formed by x-ray irradiation of a single crystal of malonic acid \cite{McConnell1960,Mehring1986}.  So far, however, this system has only been used to demonstrate two qubit gates used to prepare Bell states \cite{Mehring2003}. ENDOR techniques have also been used to demonstrate the spin-bus or $S$-bus proposal for quantum computing in $\textrm{CaF}_2$ doped with $\textrm{Ce}^{3+}$ \cite{Mehring2006} and the transfer of quantum states between electron spin and \nuc{31}{P} nuclear spin states for use as a quantum memory \cite{Morton2008}.

\subsection{Other quantum technologies}
There is a very wide range of other technologies which have been suggested for implementing quantum computing, but it is not sensible to attempt to summarise these here.  An excellent recent set of reviews has been published describing several major approaches, including trapped ions \cite{Blatt2008}, ultracold atoms trapped in optical lattices \cite{Bloch2008}, superconducting qubits \cite{Clarke2008}, spins in exotic systems such as semiconductors \cite{Hanson2008}, and possible approaches for a quantum internet \cite{Kimble2008}.

\section{Logic gates}\label{sec:gates}
Just as classical computers are built out of a small set of basic logic elements, quantum computers can be constructed from a \textit{universal set} of quantum logic gates \cite{Divincenzo1998}.  A traditional set of gates for discussing classical computation is \textsc{and}, \textsc{or} and \textsc{not}, as any desired binary logic function can be assembled from these \cite{FCbook}. In fact this set is excessive as all three of these gates can themselves be constructed as circuits of \textsc{nand} or \textsc{nor} gates, and so both \textsc{nand} and \textsc{nor} are universal for classical computation.  These two gates are also relatively simple to implement in electronic circuits.

These gates are, however, completely unsuitable for quantum computations, which proceed by a series of unitary transformations of quantum states.  Unitary transformations must be logically reversible (since they have inverses), and both \textsc{nand} and \textsc{nor} are logically irreversible, as it is not possible to reconstruct the inputs knowing only the outputs. It is possible to implement classical computations entirely reversibly \cite{FCbook}, and the three-bit \TOFFOLI\ gate, or \CCNOT\ gate, is a reversible equivalent to \NAND, and is universal for classical reversible computation.  It is not, however, universal for quantum computation, as it does not interconvert eigenstates and superposition states of qubits, and so confines qubits to a classical subspace.

Access to superposition states appears initially to create a problem for describing a universal set of quantum logic gates, as there are an infinite number of superposition states of the general form Eq.~\ref{eq:qubit}, and it is not possible to create an infinite number of different superposition states using only finite circuits built from a finite number of logic gates.  It is, however possible to find a circuit which approaches arbitrarily close to any desired unitary transformation using a number of logic gates which rises only modestly with the desired accuracy \cite{Deutsch1989,Barenco1995,Dawson2006}.  Indeed it can be shown that almost any two-qubit gate is universal in this pragmatic sense \cite{Barenco1995a,Deutsch1995,Lloyd1995}.  In practice, however, it is usually more convenient to enlarge the set of gates, and one common universal set is made up of three gates.  Two single-qubit gates, the Hadamard gate (H)
\begin{equation}\label{eq:H}
\raisebox{0.5ex}{
\Qcircuit @C=1em @R=.7em {
& \gate{\textrm{H}} & \qw
}}
\;=\;\begin{pmatrix}\frac{1}{\sqrt{2}}&\frac{1}{\sqrt{2}}\\\frac{1}{\sqrt{2}}&\frac{-1}{\sqrt{2}}\end{pmatrix}
\end{equation}
which performs the transformations
\begin{equation}
\ket{0}\overset{\text{H}}\longrightarrow\ket{+}\overset{\text{H}}\longrightarrow\ket{0}
\end{equation}
\begin{equation}
\ket{1}\overset{\text{H}}\longrightarrow\ket{-}\overset{\text{H}}\longrightarrow\ket{1}
\end{equation}
and the fourth root of Z gate (T)
\begin{equation}\label{eq:T}
\raisebox{0.5ex}{
\Qcircuit @C=1em @R=.7em {
& \gate{\textrm{T}} & \qw
}}\;=\;\begin{pmatrix}1&0\\0&\textrm{e}^{\textrm{i}\pi/4}\end{pmatrix}
\end{equation}
can be combined to make any desired single-qubit gate.  With the addition of the two-qubit \CNOT\ gate (\textsc{cnot})
\begin{equation}\label{eq:CNOT}
\raisebox{2ex}{
\Qcircuit @C=1em @R=.7em {
& \ctrl{1} &  \qw \\
& \targ & \qw}}\;=\;\begin{pmatrix}1&0&0&0\\0&1&0&0\\0&0&0&1\\0&0&1&0\end{pmatrix}.
\end{equation}
any desired gate can then be constructed with relative ease \cite{Divincenzo1998,Barenco1995}.

Important single-qubit gates include the three elementary gates X, Y, and Z, corresponding to the three Pauli matrices $\sigma_x$, $\sigma_y$, and $\sigma_z$.  Of these three gates only X is a classical logic gate, being simply a \textsc{not} gate, while Y and Z only make sense on a quantum computer.   As discussed below, however, it is usually more sensible in NMR to implement X and Y gates (and any other rotations in the $xy$-plane) using RF pulses, and then construct H and T from networks of these pulses.

\subsection{Global phases and logic gates}
Just as global phases can be ignored when considering qubit states, they can also be ignored when designing quantum logic gates, as the effect of a global phase in a logic gate is simply to impose an irrelevant global phase on any state.  Another way of looking at this is that a quantum logic gate is a propagator $U$, that will be implemented by applying some effective Hamiltonian $\mathcal{H}$ for a time $t$, such that $U=\exp(-\textrm{i}\mathcal{H}t)$.  A global phase in $U$ then corresponds to a constant offset in $\mathcal{H}$, in effect moving the zero of the energy scale.

The quantum logic gates used in NMR descriptions usually differ from their traditional forms in precisely this way, because most NMR treatments place the energy zero \textit{between} the two spin states, treating the Zeeman effect and scalar couplings as splittings around this centre, while most other implementations of quantum computing place the energy zero coincident with the lowest eigenstate.  Thus, for example, the fourth root of Z gate in NMR implementations typically takes the form
\begin{equation}\label{eq:Tnmr}
\textrm{T}=\begin{pmatrix}\textrm{e}^{-\textrm{i}\pi/8}&0\\0&\textrm{e}^{\textrm{i}\pi/8}\end{pmatrix}
\end{equation}
rather than the form above, Eq.~\ref{eq:T}.  In single qubit systems this is obviously completely irrelevant, and it is also irrelevant when a single qubit gate is applied in a system with two (or more) qubits, as the gate then takes the form $\textrm{T}\otimes\mathbf{1}$, if applied to the first qubit, or $\mathbf{1}\otimes\textrm{T}$ if applied to the second qubit.  In either case the global phase remains a global phase.

With controlled gates, however, it is necessary to be much more careful.  Consider for example the controlled-T gate, which traditionally takes the form
\begin{equation}\label{eq:cT}
%\textrm{controlled-T}=
\raisebox{2ex}{
\Qcircuit @C=1em @R=.7em {
& \ctrl{1} &  \qw \\
& \gate{\textrm{T}} & \qw}}\;=\;
\begin{pmatrix}1&0&0&0\\0&1&0&0\\0&0&1&0\\0&0&0&\textrm{e}^{\textrm{i}\pi/4}\end{pmatrix}.
\end{equation}
It might seem tempting to replace this by an NMR-like form, with a phase shift on the T-gate,
\begin{equation}\label{eq:cTwrong}
%\textrm{controlled-T}=
\begin{pmatrix}1&0&0&0\\0&1&0&0\\0&0&\textrm{e}^{-\textrm{i}\pi/8}&0\\0&0&0&\textrm{e}^{\textrm{i}\pi/8}\end{pmatrix}
\end{equation}
but this is entirely wrong, as the phase shift is now a local phase shift, and these two gates are quite different.  It is, of course, permissable to use an alternative controlled-T which differs only by a global phase, such as
\begin{equation}\label{eq:cTnmr}
%\textrm{controlled-T}=
\begin{pmatrix}\textrm{e}^{-\textrm{i}\pi/8}&0&0&0\\0&\textrm{e}^{-\textrm{i}\pi/8}&0&0\\0&0&\textrm{e}^{-\textrm{i}\pi/8}&0\\0&0&0&\textrm{e}^{\textrm{i}\pi/8}\end{pmatrix}=
\textrm{e}^{-\textrm{i}\pi/8}\begin{pmatrix}1&0&0&0\\0&1&0&0\\0&0&1&0\\0&0&0&\textrm{e}^{\textrm{i}\pi/4}\end{pmatrix}
\end{equation}
as long as the global phase shift is done correctly.

\subsection{Basic methods: single qubit gates}
As noted above, it is only necessary to implement single-qubit gates, which only change the state of a single qubit, and one non-trivial two-qubit gate.  Here non-trivial means that the final state of at least one of the two qubits involved depends on the initial states of \textit{both} qubits, so that the two-qubit gate encodes some sort of conditional logic. Non-trivial two qubit gates can interconvert product states and entangled states, and so are essential to access the full range of possible states. Trivial two qubit gates can either be written as products of single qubit gates, or as products of single qubit gates and the (trivial) two-qubit \SWAP\ gate, which simply swaps the states of two qubits \cite{Linden1999b,Madi1998}.

Single-qubit gates correspond to rotating a single spin in its own one-spin Hilbert space. For a one qubit computer, implemented using a single nuclear spin, this can be achieved by applying RF fields.  A completely general rotation can be described by three angles
\begin{equation}\label{eq:genpulse}
U(\theta,\psi,\phi)=\exp[-\textrm{i}\theta(I_x\sin\psi\cos\phi+I_y\sin\psi\sin\phi+I_z\cos\psi)]
\end{equation}
where $\psi$ describes the co-latitude of the rotation axis and product operator notation \cite{Sorensen1983} has been used. For simplicity it is often best to consider only resonant RF fields, so that $\psi=\pi/2$ and the rotation is reduced to the conventional form
\begin{equation}\label{eq:pulse}
U(\theta,\phi)=\exp[-\textrm{i}\theta(I_x\cos\phi+I_y\sin\phi)]
\end{equation}
where $\theta$ and $\phi$ are the pulse nutation and phase angles.

Pulses of this kind provide a direct route to the basic gates X and Y, which can be implemented as $180^\circ_x$ and $180^\circ_y$ rotations respectively.  Rotations about axes not in the $xy$ plane can then be implemented as sequences of pulses. Rotations around the $z$ axis are easily constructed with composite $z$-rotations \cite{Freeman1981}, using the identity
\begin{equation}\label{eq:compZ}
\theta_z=90^\circ_y\;\theta_x\;90^\circ_{-y}
\end{equation}
where the pulse sequence is written with time running from left to right, so that the leftmost pulse is the first pulse applied.  This is, of course, the reverse of the order used to describe a sequence of propagators
\begin{equation}\label{eq:compZprop}
\exp(-\textrm{i}\,\theta\,I_z)=\exp(\textrm{i}\,\pi/2\,I_y)\times\exp(-\textrm{i}\,\theta\,I_x)\times\exp(-\textrm{i}\,\pi/2\,I_y)
\end{equation}
which can lead to considerable confusion, as it is not always obvious whether a particular equation is describing a sequence of propagators or of pulses.  Some authors have tried to sidestep this by describing \textit{inverse propagators}, as these multiply in the same order as pulses are listed, but this can simply deepen the confusion.

In larger spin systems it is necessary to implement logic gates in a qubit-selective manner, so that only a single-qubit is affected by the RF field.  In a fully heteronuclear spin system qubit selection is simple, as every spin will be a long way from resonance with every other spin, and simple hard pulses applied on resonance can be used \cite{Chuang1998}, but in homonuclear systems it is essential to be more careful.

First, it is necessary to ensure that only one spin is directly affected by the pulse, which can be achieved by the use of shaped selective pulses \cite{Jones1998c,Vandersypen2001,Freeman1998}, with Gaussian \cite{Bauer1984}, Hermite \cite{Warren1984} and BURP \cite{Geen1991} pulses being particularly popular. Although many more sophisticated selective pulses are known, these must be used with great care, as they are normally optimised for particular transformations on the Bloch sphere (such as excitation or refocussing), while quantum logic gates must be implemented using general rotations, which act properly on all possible initial states. Alternatively, it is possible to use variations on ``jump and return'' and related pulse sequences \cite{Plateau1982,Hore1983} to achieve selection \cite{Jones1999a,Cummins2002,Cummins2001,Bowdrey2006}, but this is normally possible only when there are just two spins of a particular species.

Secondly, it is also important to think about the free evolution of any unexcited nuclei in the spin system.  Even if an apparently ideal selective pulse is used, any other spins will still evolve during the pulse duration at their own Larmor frequencies, picking up phases reflecting both Zeeman interactions and spin--spin couplings.  With real pulses these spins will also experience additional rotations from transient Bloch--Siegert shifts \cite{Emsley1990}, and all these phases must be considered.  Early papers only considered the effects of the direct Zeeman interaction, neglecting the two smaller terms, but correction of these (not always small) terms has become a major topic in more recent work \cite{Vandersypen2004,Ryan2008}.

The simplest approach to correcting the direct Zeeman interaction is to choose the length of the selective pulses so that the phase acquired is a multiple of $2\pi$, allowing it to be ignored \cite{Jones1998c}.  Again, this approach can be used only when there are just two spins of a particular nuclear type as the pulse sequence is synchronised with the stroboscopic alignments between the two spins occurring at time intervals of $1/\delta\nu$, where $\delta\nu$ is the difference in Larmor frequencies.  Methods based on jump and return sequences implicitly use this stroboscopic effect, and so can include the desired corrections automatically.

With more than two spins a more sophisticated approach is necessary, and the most common is the use of abstract reference frames \cite{Knill2000}.  Rather than using a single rotating frame for each nuclear species, a separate frame is used for each individual spin, with every spin being in resonance with its own frame.  It is not, of course, necessary to have a separate RF transmitter for each spin, as ``virtual transmitters'' can be created using phase ramps \cite{Freeman1998} to shift the transmitter frequency and calculating the appropriate initial phase for each spin at each point in time.  Note that when applying the same single qubit gate to two or more spins it is no longer possible to use the simple direct method of applying a hard pulse; instead it is necessary to use a sequence of selective pulses, exciting the spins one at a time, or a specially designed selective pulse addressing multiple spins.  Abstract reference frames can also be used to correct transient Bloch--Siegert shifts, by rotating each frame appropriately to absorb the additional phase shifts.  This process is conceptually straightforward but computationally tedious, and so is frequently handled by a pulse sequence compiler as described below.  Undesired evolution under spin--spin couplings can be handled in a similar way, although this process is slightly more complicated \cite{Bowdrey2005}.

Abstract reference frames simplify the implementation of single qubit gates by avoiding the necessity of correcting phase shifts ($z$-rotations), but also provide a simple way of \textit{implementing} desired $z$-rotations.  Rotating the frame instead of the spin is simply a matter of computational book keeping, and so can in principle be done instantaneously and perfectly.  For this reason it is often sensible to decompose single-qubit gates into circuits involving $z$-rotations wherever possible.  One important example is the Hadamard gate, Eq.~\ref{eq:H}, which can be implemented using the pulse sequence $90_{-y}\,180_z$ or equivalently $180_z\,90_y$.  The $z$-rotations can be absorbed into the reference frame, and so Hadamard gates can be replaced by $\textrm{h}=90_y$, sometimes called the pseudo-Hadamard gate \cite{Jones1998c,Jones1998b}, or its inverse, $\textrm{h}^{-1}=90_{-y}$.  As Hadamard gates frequently occur in pairs, these pairs can be replaced by one $\textrm{h}^{-1}$ and one $\textrm{h}$ gate, with the $180_z$ rotations cancelling out; when they do not occur in pairs it is still possible to replace them with pseudo-Hadamard gates as long as the $180_z$ rotations are absorbed into the reference frames.

An alternative approach which has grown in popularity is to use methods from optimal control theory to develop shaped pulses which both provide selective excitation and refocus undesirable interactions.  This approach is explored in more detail in Section~\ref{sec:GRAPE} below.

\subsection{Basic methods: controlled gates}
Next I turn to non-trivial two-qubit gates, and will begin by considering two-spin (two-qubit) systems.  It is in principle only necessary to construct a single gate of this kind, and the traditional gate used in most theoretical descriptions is the \CNOT\ gate, or controlled-X gate, Eq.~\ref{eq:CNOT}.  In NMR experiments, however, the key two-qubit gate is the controlled-Z gate
\begin{equation}\label{eq:cZ}
%\textrm{controlled-Z}=
\raisebox{2ex}{
\Qcircuit @C=1em @R=.7em {
& \ctrl{1} &  \qw \\
& \gate{\textrm{Z}} & \qw}}\;=\;
\begin{pmatrix}1&0&0&0\\0&1&0&0\\0&0&1&0\\0&0&0&-1\end{pmatrix}
%=\textrm{e}^{-\textrm{i}\,\pi/4}\begin{pmatrix}\textrm{e}^{\textrm{i}\,\pi/4}&0&0&0\\0&\textrm{e}^{\textrm{i}\,\pi/4}&0&0\\0&0&\textrm{e}^{\textrm{i}\,\pi/4}&0\\0&0&0&\textrm{e}^{-\textrm{i}\,3\pi/4}\end{pmatrix}
\end{equation}
which is easily converted to a \CNOT\ gate
\begin{equation}
\mbox{
\Qcircuit @C=1em @R=.7em {
& \ctrl{1} &  \qw \\
& \targ & \qw
}}
\;\:\raisebox{-2ex}{$=$}\;\:
\mbox{
\Qcircuit @C=1em @R=.7em {
& \ctrl{1} &  \qw \\
& \gate{\textrm{X}} & \qw
}}
\;\:\raisebox{-2ex}{$=$}\;\:
\mbox{
\Qcircuit @C=1em @R=.7em {
& \qw & \ctrl{1} &  \qw & \qw\\
& \gate{\textrm{H}} &\gate{\textrm{Z}} & \gate{\textrm{H}} & \qw
}}
\end{equation}
with the application of a pair of Hadamard gates \cite{Jones2001a}.  Note that, just like pulse sequences, gate networks are written with time running from left to right.

Unlike the \CNOT\ gate, controlled-Z is symmetric between the two spins, which is the form expected for a spin--spin interaction, and it can be easily decomposed with product operators \cite{Sorensen1983}
\begin{align}\label{eq:cZpo}
\textrm{controlled-Z}&=\exp[-\textrm{i}\,(\pi/2)\,(\half{E}-I_z-S_z+2I_zS_z)]\\
                     &=\exp[-\textrm{i}\,(\pi/2)\,(-\half{E}+I_z+S_z-2I_zS_z)]
\end{align}
where the choice between the two decompositions is simply a matter of convenience.  All four terms commute, and so can be considered individually.  The $\half{E}$ term is just a global phase, and can be ignored as usual.  Terms in $I_z$ and $S_z$ are just single qubit rotations, and can be implemented with single-qubit gates, or simply absorbed into the reference frame.  This leaves the only important term, which in Eq.~\ref{eq:cZpo} corresponds to evolution under the (Ising like) spin--spin coupling term, $\pi{J}\,2I_zS_z$ for a time $1/2J$; note that there is a close analogy between controlled logic gates and spin-state-selective coherence transfer sequences \cite{Meissner1997,Meissner1997a} in conventional NMR.  The spin Hamiltonian will include both Zeeman and coupling terms, but conventional spin-echo sequences can be used to remove the undesirable terms \cite{Linden1999a}.  Similar methods can, of course, be used to generate closely related gates \cite{Jones1998b,Jones2001a}, such as the controlled-T gate, Eq.~\ref{eq:cT}.

Similar methods can also be used in systems with more than two spins; the key element lies in the use of spin-echo sequences to isolate the desired spin--spin coupling, while suppressing all undesired interactions.  Conventional approaches, based on nesting spin-echoes, become very complicated in large fully-coupled spin systems, and this should instead be done efficiently using sequences based on Hadamard matrices \cite{Linden1999,Jones1999,Leung2000}.  In a system of $n$ spins, all of which are coupled weakly to one another, any desired coupling can be isolated with around $n^2$ pulses, although it is not always necessary to refocus all the additional couplings \cite{Bowdrey2005}.

Most spin systems are not fully coupled, and a much smaller number of pulses can be used to isolate desired couplings in such systems \cite{Linden1999,Jones1999}.  If a desired coupling does not exist, or is too small to be useful, then it must be generated indirectly.  This is straightforward in principle, as quantum \SWAP\ gates \cite{Linden1999b,Madi1998} can be used to move logical qubits around a spin system to bring them into contact; in practice it is not necessary to use full \SWAP\ gates, and considerably more efficient approaches are known \cite{Collins2000,Khaneja2007}.  It is, of course, essential that every pair of spins must be connected either directly or indirectly by some chain of usable couplings: it is not possible to perform two-qubit gates between spins in completely separate spin systems.

More complex gates, such as the \TOFFOLI\ gate, or controlled-controlled-\NOT, can then be constructed using quantum circuits \cite{Divincenzo1998,Barenco1995}, but it is often better to construct these gates directly from NMR primitives, as this can lead to significantly simpler implementations \cite{Price1999,Price2000}.

\subsection{Transition selective pulses}
An alternative, conceptually simpler, approach to implementing controlled gates is simply to use transition selective pulses \cite{Barenco1995b}.  The NMR spectrum of a two-spin system will contain two doublets, and a sufficiently long selective pulse will only excite one transition in a doublet, and so will only affect the state of the excited spin if the coupled partner is in the desired state.  It might seem that applying a $180^\circ_x$ pulse to one transition would implement a \CNOT\ gate, but it is necessary to think carefully about global phases.  The propagator corresponding to a $180^\circ_x$ rotation is $-\textrm{i}\,\sigma_x$, which differs from a \NOT\ gate by a global phase, reflecting spinor behaviour \cite{Cory1998a}.  For a transition selective pulse \cite{Havel2004} this phase is no longer global, and so cannot be ignored: the propagator is
\begin{equation}\label{eq:CNOTts}
\begin{pmatrix}1&0&0&0\\0&1&0&0\\0&0&0&-\textrm{i}\\0&0&-\textrm{i}&0\end{pmatrix}
\end{equation}
which differs from a \CNOT\ gate, Eq.~\ref{eq:CNOT}, by a $z$-rotation on the control spin.

In a two-spin system this approach has no particular advantage, and is indeed fundamentally equivalent to the more traditional approach based on pulses and delays, which can be considered as a ``jump and return'' style implementation \cite{Bowdrey2006} of a transition selective pulse.  Alternatively, the transition selective pulse method relates to the pulses and delays method much as the selective population transfer experiment \cite{Pachler1973} relates to INEPT \cite{Morris1979}.  In multi-spin systems, however, transition selective pulses provide a simple and direct way of implementing more complex quantum logic gates.  For example, in a three-spin system the \TOFFOLI\ gate can be implemented directly as a transition selective pulse addressing one of the four lines in the double doublet corresponding to the target spin \cite{Dorai2000a,Mahesh2001,Linden1998,Du2000a,Du2001a}.  A corresponding disadvantage of this method is, of course, that constructing a two qubit gate in this system will require a pulse which selects two of the four transitions for the target spin \cite{Jones2001a}.

\subsection{Composite pulses}
Composite pulses \cite{Levitt1986} have found widespread use in conventional NMR experiments to reduce the effects of a wide range of experimental imperfections, most notably off-resonance effects and pulse length errors arising from RF inhomogeneity.  Similar imperfections are likely to affect most experimental implementations of quantum information processing and there has been considerable interest in applying these ideas.  Most conventional composite pulses are not, however, suitable for use in quantum computers, as they are optimised for particular classes of initial state: for example, most composite $180^\circ$ pulses are optimised either for inverting the population of a spin, or for producing a spin echo on a coherent superposition.  By contrast, pulses used on quantum computers must be general rotors, which perform well for any initial state. Composite pulses of this kind are rarely used in conventional NMR, but a small number of so-called Class~A composite pulses \cite{Levitt1986} are known, and these have been developed for wider use \cite{Jones2001a,Jones2003a,Cummins2003}.  A method for constructing general rotors from conventional point-to-point pulses has also been described \cite{Luy2005}.

An early composite $90^\circ$ pulse tackling off-resonance errors was described by Tycko \cite{Tycko1985}, replacing a $90^\circ_x$ pulse with the three pulse sequence $385^\circ_x320^\circ_{-x}25^\circ_x$, has been generalised to give the \textsc{corpse} family of composite pulses \cite{Cummins2003,Cummins2000}, in which a $\theta_x$ pulse is replaced by three pulses, applied along the $+x$, $-x$ and $+x$ axes as before, with flip angles given by
\begin{align}
\theta_1&=2n_1\pi+\frac{\theta}{2}-\arcsin\left(\frac{\sin(\theta/2)}{2}\right)\\
\theta_2&=2n_2\pi-2\arcsin\left(\frac{\sin(\theta/2)}{2}\right)\\
\theta_3&=2n_3\pi+\frac{\theta}{2}-\arcsin\left(\frac{\sin(\theta/2)}{2}\right)
\end{align}
where $n_1$, $n_2$ and $n_3$ are integers, with the best results  \cite{Cummins2003} occurring for $n_1=n_2=1$ and $n_3=0$.  These sequences have been demonstrated by NMR \cite{Cummins2000} and \textsc{squid} \cite{Collin2004} experiments.  Pulse sequences have also been designed which are tailored for particular off-resonance effects \cite{Cummins2001}.

There has been considerably more interest in composite pulses to tackle pulse length errors, which can largely be traced back to a three pulse composite $180^\circ$ pulse due to Tycko and coworkers \cite{Tycko1985} or to the BB1 family of sequences discovered by Wimperis \cite{Wimperis1994}.  Tycko's pulse sequence has been generalised to give the \textsc{scrofulous} family of composite pulses \cite{Cummins2003}, but using the BB1 family is preferable in most cases.

BB1 differs from many other composite pulses in that it seeks to design an error-correcting pulse, which can be combined with the naive error-prone pulse to give a more accurate compound pulse, much as a contact lens can be used to correct eyesight.  Originally \cite{Wimperis1994} this error correcting sequence (sometimes called a W1 sequence) was placed before the naive pulse, but it can instead be placed after the naive pulse, or indeed in the middle of it \cite{Cummins2003,Xiao2006}.  It comprises three pulses, in the form $180^\circ_{\phi_1}\,360^\circ_{\phi_2}\,180^\circ_{\phi_1}$, with $\phi_2=3\,\phi_1$ and
\begin{equation}
\phi_1=\pm\arccos\left(-\frac{\theta}{4\pi}\right)
\end{equation}
where the choice of sign is unimportant as long as it is made consistently.  The method can be extended to build two-qubit gates which are robust to variations in the size of the underlying scalar coupling \cite{Jones2003b,Jones2003,Xiao2006}

BB1 has proved a remarkably successful composite pulse, and is surprisingly difficult to improve upon.  The quality of a composite pulse for quantum computing can be assessed in various ways, but in practice there are two important families of approaches.  The most direct approach is to expand the propagator for the composite pulse as a power series in the size of the error, and determine the size and order of the lowest order error term.  As an example consider implementing a $180^\circ_x$ pulse using a naive pulse with a fractional pulse length error of $\epsilon$, so that the flip angle of the pulse is in fact $180^\circ\times(1+\epsilon)$.  The propagator is then
\begin{equation}
\exp[-\textrm{i}\,\pi(1+\epsilon)I_x]=
\begin{pmatrix}0&-\textrm{i}\\-\textrm{i}&0\end{pmatrix}
-\epsilon\begin{pmatrix}\pi/2&0\\0&\pi/2\end{pmatrix}+\textrm{O}(\epsilon^2)
\end{equation}
and so the naive pulse has an error of order $\epsilon$.  Alternatively the quality can be assessed by calculating the \textit{propagator fidelity} $F$ between the desired propagator $U$ and the actual propagator $V$, given by
\begin{equation}
F=|\textrm{Tr}(VU^{-1})/\textrm{Tr}(UU^{-1})|,
\end{equation}
and then expanding the fidelity as a power series in the error.  For the naive pulse considered above the fidelity is
\begin{equation}
F=1-\epsilon^2\pi^2/8+\textrm{O}(\epsilon^4)
\end{equation}
and so the naive pulse has \textit{infidelity} of order $\epsilon^2$.  The difference between these two methods of assessing a pulse must be borne in mind when comparing pulses in different papers; in general an error of order $n$ will correspond to an infidelity of order $2n$.

BB1 pulses can be derived by designing composite pulses which suppress first order pulse length errors, but it turns out that BB1 also suppresses second order errors automatically, leaving only third order errors (sixth order infidelity).  It is not clear why this fortuitous double cancellation occurs; it is not a general feature of composite pulses.  Other pulses with similar properties are known \cite{McHugh2005}, but these have no advantages over BB1.  Beyond this, BB1 pulses are also relatively robust to off resonance-errors \cite{Cummins2003}, and generally insensitive to small errors in their implementation, so that BB1 pulses work in practice very much as expected from theory \cite{Xiao2006}.  In addition to NMR experiments \cite{Xiao2005,Xiao2006} BB1 pulses have been demonstrated in electron spin resonance \cite{Morton2005a,Ardavan2007a} and have inspired applications in other fields \cite{Wesenberg2003,Ardavan2007b,Testolin2007,Timoney2008}.

Although BB1 has proved highly successful, it is obviously interesting to seek still better pulse sequences, and Brown \textit{et al.} have tackled this in two ways \cite{Brown2004,Brown2005}.  Firstly they have shown how the BB1 approach can, in effect, be nested, creating ever higher orders of simultaneous correction.  A robust $90^\circ$ pulse from the B4 family of pulses (which remove the third order error term) has been implemented in NMR experiments \cite{Xiao2006}, but this composite pulse is very long (the correction sequence contains 27 pulses with a total length equivalent to a $7200^\circ$ rotation) and does not perform much better than BB1.  Secondly they have described a general method, using insights from the Solovay--Kitaev theorem \cite{Dawson2006}, showing how arbitrarily accurate composite pulses can be constructed in general by building a series of correction sequences which correct errors one order at a time.  An expanded version of part of their method written in more conventional NMR notation is also available \cite{Alway2007}.  These ideas have subsequently been extended to multi-qubit systems \cite{Tomita2010}.

\subsection{Strongly modulated composite pulses}\label{sec:SMCP}
Strongly modulated composite pulses \cite{Fortunato2002,Boulant2003,Weinstein2004} are not composite pulses in the conventional sense, but are in many ways closer to shaped pulses \cite{Freeman1998}.  Conventional composite pulses are comprised of a small number of pulses which usually vary only in length and phase, while shaped pulses are made up from a large number of pulses of the same length, with control of both amplitude and phase.  Strongly modulated composite pulses contain a small number of pulses, but these pulses can differ in amplitude and \textit{frequency}, as well as length and phase.  In practice strongly modulated pulses are always implemented as shaped pulses, with frequency shifts implemented as phase ramps \cite{Freeman1998}, so that strong modulation can be thought of as an unusual method for parameterising a shaped pulse, and strongly modulated pulses are designed by numerical simulation rather than analytical calculation.  Unlike most conventional methods, however, the propagator for the pulse can be calculated using the short composite pulse description, rather than the much longer shaped pulse description, simplifying calculations.  It is, of course, necessary to ensure that the sufficient digitisation is used when implementing the composite pulse as a shaped pulse.

Strongly modulated pulses were originally developed to allow desired quantum logic gates to be implemented in complex multi-spin systems.  The conventional approach, using spin-echoes to isolate desired terms in the spin Hamiltonian and then constructing a logic circuit from these elementary terms, can become impractical in larger spin systems, due both to the large number of interactions which have to be considered and interactions between non-idealities arising from individual pulses.  Strongly modulated pulses seek to implement the desired logic gate in one go, simply optimising the propagator of the whole composite pulse, calculated from the sum of the background and RF Hamiltonians, by changing the parameters describing the RF component \cite{Fortunato2002}.  This also allows the pulses to correct for any non-idealities in the system: for example, strongly modulated pulses can deal with the slight deviations from weak coupling found in homonuclear systems, by simply using the full coupling term in the Hamiltonian used to calculate the pulse propagator.  In the same way non-idealities, such as transient Bloch--Siegert shifts and interactions between selective pulses, are automatically incorporated in the calculation and corrected for in the pulse sequence, as can effects such as short delays around pulses imposed by the spectrometer hardware.  The effects of incoherent effects, such as RF inhomogeneity, are not automatically incorporated, but can be included by simultaneously optimising the propagator for a range of RF strengths \cite{Fortunato2002,Pravia2003}; this also allows incoherent effects to be distinguished from decoherent effects \cite{Weinstein2004}.

The discussion above assumes that a strongly modulated pulse can be found to implement a desired propagator.  As the NMR Hamiltonian supplemented by RF fields provides a universal set of quantum controls, it should always be possible (neglecting RF inhomogeneity) to find a suitable pulse, as long as a sufficient number of pulses and total pulse length are available.  The required number of pulses is usually found by trial and error, with the number of pulses being increased until a suitable solution is found.  The total pulse length required can be estimated from optimal control theory (see below), and then optimised by the algorithm, or can simply be found by trial and error.  It is usual to start optimisation from a fairly smooth low power initial pulse, and to use several initial pulse sequences, retaining those which seem to converge to an acceptable solution for further optimisation.  If desired penalty functions can be added to dissuade the algorithm from designing pulses with excessive length, offset frequencies, or RF power, but this is not always necessary.  In practice it is usually possible to find suitable pulse sequences fairly easily, which perform desired transformations with high fidelity (values above 0.999 are usually possible).  When the effects of RF inhomogeneity are included the process becomes a little more complicated, but it is often possible to find sequences with fidelities above 0.99 over a reasonable range of inhomogeneities.

Strongly modulated pulses have been demonstrated in a range of NMR quantum information processing experiments with spin-$\half$ nuclei in the liquid state \cite{Negrevergne2006,Fortunato2002,Fitzsimons2007,Du2007,Mitra2008,Boulant2003,Weinstein2004,Pravia2003,Cappellaro2005,Anwar2005,Hodges2007} or solid state \cite{Baugh2005,Baugh2006}, as well as to strongly coupled systems \cite{Mahesh2006}, quadrupolar nuclei \cite{Teles2007,Kampermann2005} and ENDOR \cite{Hodges2008}.  Until the arrival of GRAPE techniques, described below, it seemed likely that strong modulation would become the dominant method for designing pulses for NMR quantum information processing.  There is, however, some concern as to whether using such pulses is in some sense cheating, as their design requires a full simulation of the entire spin system, and the difficulty of the simulation increases exponentially with the number of spins involved  \cite{Negrevergne2006}.  Indeed, the best way to design strongly modulated pulses would be to use a quantum computer!  For this reason some authors have avoided strongly modulated pulses where possible, and have selected spin systems carefully to minimise these problems \cite{Ryan2005}.

\subsection{GRAPE and other methods from optimal control}\label{sec:GRAPE}
As strong modulation is simply an unusual method of parameterising a shaped pulse, allowing the numerical optimisation of arbitrary transformations with arbitrary Hamiltonians, one might wonder why shaped pulses shouldn't be designed by brute force numerical optimisation of the amplitude and phase of each period, with no attempt at imposing a particular parameterisation.  Historically this has not been done, due to a perception that optimising over a large number (hundreds or even thousands) of amplitudes and phases would be computationally infeasible.

Despite these perceived difficulties there has always been some interest in this approach, arising from the field of optimal control theory.  Once it has been established whether a transformation is possible in  principle \cite{Glaser1998}, the next obvious question is how the required resources, such as the time required to implement the transformation, can be minimised.  While there has been some abstract work on this question from the viewpoint of quantum computation \cite{Vidal2002,Childs2003,Zeier2004} a more profitable approach for pulse design is provided by the field of coherent control \cite{Grace2007,Rabitz2009,Brif2010}.  Early applications to NMR largely studied methods for performing coherence transfers between spins in the minimum possible time \cite{Khaneja2001,Reiss2002,Khaneja2005a}, including relayed transfers via intermediate spins \cite{Khaneja2002}; these ideas have obvious applications in branches of NMR spectroscopy where signal loss due to spin--spin relaxation during coherence transfers is a problem \cite{Khaneja2003,Stefanatos2005,Gershenzon2007}.

These ideas were then generalised to the implementation of quantum logic gates \cite{Schulte-Herbruggen2005} and the \GRAPE\ (gradient ascent pulse engineering) algorithm was described \cite{Khaneja2005} which allows arbitrary shaped pulses to be developed to perform desired unitary transformations by adding control fields to a background Hamiltonian.  The key idea behind \GRAPE\ is a simple and efficient method \cite{Khaneja2005} for estimating the gradient of the fidelity between a shaped pulse and the desired transformation, thus allowing conventional optimisation methods to be used to design pulses.  The traditional method for estimating a gradient for a composite pulse with $n$ independent time periods, and so described by $2n$ variables, requires $2n+1$ complete evaluations of the pulse propagator, but the \GRAPE\ approach enables the gradient to be estimated using only $2$ complete evaluations, by eliminating unnecessarily repeated calculations.  This makes it possible to optimise shaped pulses described by very large numbers of parameters (hundreds, or even thousands of time periods).

\GRAPE\ methods have been used to design a wide variety of exotic shaped pulses with applications in conventional NMR \cite{Kobzar2005,Vosegaard2005,Skinner2006,Kehlet2007} and to quantum information processing using both NMR \cite{Ryan2008a,Mottonen2006,Henry2007a} and other techniques \cite{Timoney2008,Hodges2008,Sporl2007}.  An excellent recent description of the approach can be found in \cite{Ryan2008}.  Two particularly interesting developments are \GRAPE\ pulses which work entirely by phase modulation \cite{Skinner2006}, and give compensation for RF inhomogeneity over extremely wide ranges, and pulses which work entirely by amplitude modulation (except for $180^\circ$ phase shifts) \cite{Hodges2008} and so can be used in experimental fields where small angle phase shifts are not available.

Given that \GRAPE\ pulses are in principle more powerful than strongly modulated composite pulses, one might expect them to become the method of choice in NMR quantum information processing, and there are some signs of this happening.  However, \GRAPE\ pulses are vulnerable to the same fundamental criticism as strongly modulated composite pulses, namely that their design requires a full simulation of the entire spin system.  An attempt has been made to mitigate this problem by combining \GRAPE\ methods with pulse sequence compilers, as described below.

\subsection{Dynamical decoupling}
Dynamical decoupling \cite{Viola1999,Viola2000,Viola1999a} extends the ideas of coherent control to the problem of controlling open quantum systems, such as quantum systems undergoing decoherence.  The ideas can be related to ideas in conventional NMR \cite{Viola1999a,Viola2002}, including the use of decoupling techniques to suppress unwanted couplings, and the use of spin locking to create an effective relaxation time $\textrm{T}_{1\rho}$. These ideas have been extensively studied \cite{Khodjasteh2005,Cappellaro2006,Khodjasteh2007,Uhrig2007,Uhrig2009} and have found applications in NMR \cite{Krojanski2006} and EPR \cite{Morton2006a}.  Conceptual links can also be drawn between dynamical decoupling and the use of decoherence free subspaces, discussed in Section~\ref{sec:ECDFS} below.

\subsection{Pulse sequence compilers}
The pulse sequences used in NMR quantum information processing are, even by the standards of NMR, extremely long and complicated.  For example, one pulse sequence used in an NMR implementation of Shor's quantum factoring algorithm contains 299 shaped pulses \cite{Vandersypen2001}.  While some solid state NMR experiments can involve larger numbers of pulses, these are usually found in highly repetitive decoupling or mixing sequences, rather than individually placed pulses.  The sheer complexity of writing such pulse sequences has led many groups to develop simple computerised tools, usually called pulse sequence compilers, which can handle many of the more mundane steps in the calculation.

Pulse sequence compilers can act at a variety of levels, from the abstract network description, where they can be used to cancel unnecessary gates \cite{Vandersypen2001}, to low level frame tracking, where they are used to keep a record of frame rotations and extraneous couplings \cite{Ryan2008,Bowdrey2005}.  The combined use of \GRAPE\ pulses, designed to work within a spin subspace, and pulse compilers which keep track of interactions with spins outside the subspace seems particularly powerful, and is described in detail elsewhere \cite{Ryan2008}.

\subsection{Geometric phase gates}
Geometric phases are a topic of considerable interest in many branches of physics \cite{SWbook,Berry1984,Zwanziger1990a,Anandan1992,Sjoqvist2000}, and have over many years been explored in the context of NMR \cite{Suter1987,Suter1988,Skrynnikov1994,Goldman1996}, ESR \cite{Gamliel1989,Goldman1998} and NQR \cite{Tycko1987,Zee1988,Zwanziger1990,Haerle1993,Appelt1994,Appelt1995,Jones1995,Jones1997a}.  The simplest example of a geometric phase is Berry's phase \cite{Berry1984}, which arises when a Hamiltonian is adiabatically varied around a circuit.  A quantum system which starts in an eigenstate of the initial Hamiltonian will always remain in the corresponding instantaneous eigenstate of the time-varying Hamiltonian (as the process is adiabatic) and so will return to its initial state (as the process is cyclic).  The state can, of course, acquire a phase in this process, and as well as the conventional dynamic phase, which depends on the average energy of the state and the time taken for the process, the state will acquire an additional geometric phase, whose size depends only on the solid angle subtended by the circuit in some appropriate parameter space, and in particular is independent of the time taken to complete the circuit.

A simple way of thinking about these geometric phases is that the motion of the Hamiltonian leads to a fictitious magnetic field (a gauge field) \cite{Tycko1987} which interacts with the spins.  The size of the field is proportional to the rate of change of the Hamiltonian, and so the product of the field strength and the cycle time is constant.  This picture makes clear that different spin states acquire different geometric phases, so that the Berry phase is not a global phase, and the difference between two geometric phases can be detected as a relative phase shift in the evolution of a superposition.  Alternatively, geometric phases can be detected in interference experiments; in effect these work by replacing the global phase by a controlled phase, controlled by the ``which way'' path information \cite{Suter1988}.

Berry's phase can be generalised in several important ways.  In NQR experiments the degeneracy of spin states means that the system need not return to the same state at the end of a cyclic evolution; instead mixing can occur within the degenerate spin states.  Unlike phase shifts, this more general mixing need not be Abelian \cite{Zee1988}, which can lead to more complex behaviour \cite{Zwanziger1990} than that seen for spin-$\half$ nuclei. More usefully, Aharonov and Anandan have shown \cite{Aharonov1987} that Berry's phase can be seen as the adiabatic limit of a much more general phase, and that in non-adiabatic cases this phase depends on the trajectory of the state, not the trajectory of the Hamiltonian.  Such Aharonov--Anandan phases are relatively simple to observe \cite{Suter1988}; indeed they frequently arise in quite conventional NMR experiments.  For example, the fact that evolution under two $180^\circ$ pulses with phases differing by $\phi$ is equivalent (up to a global phase) to a $z$-rotation through an angle of $2\phi$ can be understood as an Aharonov--Anandan phase \cite{Suter1988}.

In addition to their fundamental interest, it has been widely suggested that geometric phases could be technically important, as their size depends only on the geometry of the cyclic path and not on other details, and thus geometric phases might be robust to variations in experimental parameters.  It is, however, not clear that such robustness can always be achieved.  Firstly, the geometric phase usually (but not always \cite{Unanyan2004}) occurs on top of a (usually much larger) dynamic phase, and it is necessary to remove this dynamic phase term.  This can usually be achieved by refocussing it with a spin-echo sequence, but the quality of the geometric phase will then depend on the quality of the refocussing step.  Secondly, the geometry of the path can itself depend on experimental parameters, and it is necessary to think carefully about exactly what determines the geometry in a particular physical situation.  In particular, with Aharonov--Anandan phases it is important to ensure that the state evolution does in fact form a closed cycle; if it does not then the path must be closed with a geodesic, changing the solid angle subtended by the path from its naive value.

Despite these caveats there has been considerable interest in the use of geometric phases to implement quantum logic gates.  The first experimental demonstration \cite{Jones2000b,Ekert2000} used the controlled-acquisition of Berry phases to implement a two-qubit quantum logic gate in the heteronuclear two-spin system provided by \nuc{13}{C} labelled chloroform.  This has been followed by many experiments and proposed experiments in NMR \cite{Gopinath2008,Wang2001,Zhu2003,Du2003,Das2005,Gopinath2006b,Du2006,Chen2007a} and ESR \cite{Morton2006,Morton2006a}, and in other systems \cite{Falci2000,Duan2001,Lloyd2001,Recati2002,Garcia-Ripoll2003a,Leibfried2003,Leek2007} (many of these later papers are largely based on an earlier proposal for holonomic quantum computation \cite{Zanardi1999}, which uses non-Abelian phase mixing effects).  There have also been theoretical studies of the robustness of such gates against decoherence \cite{Nazir2002}, variations in control parameters \cite{Blais2003,DeChiara2003,Zhu2005,Hou2007}, and non-cyclic evolution \cite{Friedenauer2003}.

\subsection{Quantum cellular automata}
The discussion so far has assumed that it is necessary to selectively address single qubits and pairs of qubits in order to implement general quantum computations.  In fact it has been known for many years that quantum information processing can be implemented in systems with much less control of individual qubits \cite{Lloyd1993,Lloyd1994,Benjamin1999,Benjamin2000,Benjamin2001,Levy2002,Benjamin2002,Twamley2003,Raussendorf2005,Raussendorf2005a,Fitzsimons2006}, sometimes referred to as \textit{quantum cellular automata}, or QCA, although that this term is also used to describe classical cellular automata implemented with quantum dots or other quantum technologies \cite{Grossing1988,Lent1993,Biafore1994,Orlov1997,Benjamin1997,Cowburn2000,Lent2003}, and, to add further confusion, there is occasional discussion of implementing quantum information processing on quantum dot cellular automata, sometimes called coherent quantum dot cellular automata \cite{Toth2001}.  In this discussion I will only consider schemes capable of implementing quantum information processing without selective qubit addressing; these ideas may have important applications to quantum information processing in optical lattices \cite{Joo2006}.

Quantum cellular automata proceed by dividing up spins into two or more groups (usually labelled A, B and so on), such that all the spins of a particular type can be addressed simultaneously but it is not possible to separately address individual spins within a given type.  The spins are assumed to interact through some network of couplings, typically acting between nearest neighbours in a linear chain or two dimensional array.  The simplest practical scheme \cite{Fitzsimons2007,Fitzsimons2006} assumes a linear chain of spins, all of the same type and connected by Ising couplings.  In systems of this kind it is always possible to address the end spins separately from the spins inside the chain, and this provides sufficient control to implement full quantum computing \cite{Fitzsimons2007}; it is not necessary to be able to distinguish the two end spins from each other, although this does help.  It is even theoretically possible to use a completely symmetric ring of spins \cite{Raussendorf2005a}, but it is not clear that this scheme can in fact be implemented in physical systems.

The simple linear chain scheme has been implemented on an NMR quantum cellular automaton using three \nuc{13}{C} nuclear spins in alanine \cite{Fitzsimons2007}, which was used to implement both a two-qubit and a three-qubit algorithm.  The system was also used to explore quantum mirroring, in which qubits are moved along a chain to their mirror-image locations.  As mirroring moves a qubit from one end of a chain to the other it can be used to transfer spin states along a quantum wire, thus moving a qubit around a quantum computer \cite{Khaneja2002a,Christandl2004,Yung2006,Cappellaro2007}.

In addition to quantum mirroring, there has also been interest in implementing particular quantum information processing tasks which do not require complete control of a quantum system; indeed, in some cases the implementation may be simpler if the spin system has a QCA structure.  One obvious and important example of this is entanglement assisted spin state measurement \cite{Cappellaro2005,Perez-Delgado2006}, and the related technique of entanglement assisted magnetic field sensing \cite{Jones2009,Simmons2010}, where the use of a star-topology QCA permits a highly entangled state to be prepared with relative ease.

\subsection{Quadrupolar nuclei and strongly coupled systems}
The methods used to implement quantum logic gates in quadrupolar systems are not fundamentally different from those used for spin-$\half$ nuclei, but there is a difference of emphasis.  The large size of quadrupolar couplings means that transition-selective pulses are simple to apply, and provide the most natural way of implementing \CNOT\ gates \cite{Khitrin2000,Sarthour2003,Sinha2001}; they can also be used to implement controlled phase-shift gates using geometric phases \cite{Gopinath2008}.  Much is often made of the relative simplicity of implementing two-qubit gates, especially the \SWAP\ gate \cite{Sinha2001} in such systems, but it must be remembered that implementing single-qubit gates can be correspondingly difficult, requiring pulses which simultaneously or sequentially select multiple transitions.  It is also necessary to consider carefully the effects of background evolution under the quadrupolar coupling on spin states not affected by selective pulses \cite{Bonk2005}.

The methods used to implement quantum logic gates in strongly coupled systems are frequently complicated and depend strongly on the details of the system.  In spin-systems where the weak coupling approximation is starting to break down the use of strongly modulated composite pulses \cite{Baugh2005,Baugh2006} or \GRAPE\ pulses \cite{Ryan2008a} provides a simple solution.  In more strongly coupled systems strongly modulated composite pulses \cite{Mahesh2006} have been used, but the use of single transition selective pulses \cite{Das2004b} or methods adapted from solid state NMR \cite{Lee2007} may be more appropriate.

\section{Initialisation}\label{sec:initialisation}
Initialisation is the process of preparing qubits in some well defined initial state from which a computation can proceed.  This process is not much discussed in conventional computation, as it is seen as trivial and uncontroversial, but it plays an important and explicit role in reversible computation \cite{FCbook}, as the initialisation step is fundamentally irreversible (the bits must be set to the desired initial state whatever their state was before the initialisation process).  This fact leads to a minimum energy cost for computation \cite{Landauer1961}, resolving the apparent paradox of Maxwell's Demon.

The situation is perhaps even more stark for quantum computers, where Mermin has forcefully argued \cite{Merminbook,Mermin2006} that the state of a qubit is not even a meaningful concept until the qubit has been initialised by a measurement.  Whether or not one accepts Mermin's arguments, it is clear that initialisation and measurement are intimately connected.  The simplest way of initialising a qubit is to measure it in the computational basis: if the result is \ket{0} then the qubit should be left alone, while if the result is \ket{1} then it should be flipped with a \NOT\ gate.  This procedure, however, requires access to a true quantum measurement device, which projects the qubit into the measurement basis.  As we shall see in the discussion on readout, conventional NMR measurements do not have this property, and so cannot be used to initialise spin states.  Instead it is necessary to use indirect methods, in which the spin state is implicitly measured by the environment.
%As the aim of this process is to prepare a spin system in a single well defined quantum state, so that

\subsection{Preparing pure states by cooling}
The most obvious method for preparing a spin system (or indeed any system) into a pure quantum state is by cooling it into its ground state.  Considering the Boltzmann expression it is immediately clear that this will only work if the thermal energy $kT$ is much less than the energy difference $\Delta{E}=h\nu$ between the ground and first excited state.  For NMR with a Larmor frequency below 1\,GHz this requires $T\ll0.05\,\textrm{K}$, which is clearly impractical for most liquid state samples, although it can be reached by solid state NMR or for liquid state studies of \nuc{3}{He} \cite{Bird2003}.  For this reason, most implementations of NMR quantum information processing have not used pure states, but rather pseudo-pure states as described below.  There is also considerable interest in developing methods for generating non-Boltzmann populations in spin systems \cite{Jones2000a}, although with the exception of techniques based on para-hydrogen \cite{Hubler2000,Anwar2004,Bowers1986,Natterer1997,Duckett1999} these attempts have not yet been completely successful.  The situation is, of course, more promising for electron spins, due to the much higher Larmor frequencies involved, although cooling to the solid state is still required \cite{Takahashi2008}.

One important point that is frequently neglected is that it is not really sufficient to have just a method for preparing a pure state at the start of a computation.  Instead, many quantum information processing methods, most notably error correction, require the ability to reinitialise at least some qubits in the middle of a computation. For this reason, no method yet demonstrated is entirely satisfactory, although some methods could in principle be extended to permit this.

\subsection{Pseudo-pure states and entanglement}
Because of the difficulty in preparing pure spin states in NMR systems, almost all NMR quantum information processing experiments have used \textit{pseudo-pure states} \cite{Cory1996,Cory1997,Cory1998a}, sometimes called \textit{effective pure states} \cite{Gershenfeld1997}.  There are many different ways of preparing such states, which will be discussed below, but it is useful to begin by considering their general properties.

A pseudo-pure state in a system of $n$ spins is simply a mixed state of the form
\begin{equation}\label{eq:ppure}
(1-\epsilon)\frac{\mathbf{1}}{2^n}+\epsilon\ket{\psi}\bra{\psi}
\end{equation}
where \ket{\psi} is the corresponding pure state, $\mathbf{1}/2^n$ is the \textit{maximally mixed state} (that is, equal probability of the system being found in any particular state), and $\epsilon$, the excess probability of finding the system in the desired pure state, can be taken as a measure of the purity of the state.  Since quantum computations proceed through a series of unitary transformations the pure component of a pseudo-pure state will evolve in exactly the same way as a pure state would, while the maximally mixed state is unaffected by any unitary transformation; thus the behaviour of a pseudo-pure state is effectively the same as that of a pure state.  For initialisation the pure component $\ket{\psi}$ can be taken to be the ground state $\ket{\textbf{0}}=\ket{00\dots0}$.

Significant differences will, of course, be seen at the end of the computation, when an attempt is made to measure the final state of the quantum computer.  With conventional quantum measurements the result may correspond to some component from the maximally mixed state, rather than the desired pure component, and for the low purities found in NMR systems such results might be expected to overwhelm the tiny pure component.  However, NMR measurements are not sensitive to the maximally mixed state (in effect, the signals from the different components of the maximally mixed state cancel out), and only the signal from the pure component is observed.  Of course the size of the observable signal \textit{does} depend on the purity, and so pseudo-pure states will give much smaller signals than pure states.

This small signal size might not immediately concern NMR spectroscopists who are accustomed to observing signals from spin systems with very low polarisations.  It is, however, important to realise that the states observed in most conventional NMR spectra are not pseudo-pure states \cite{Jones1998b}, and so instinct may not be reliable.  This problem was immediately raised by Warren \cite{Warren1997}, who showed that the maximum pseudo-pure state signal obtainable from a system of $n$ identical spins in thermal equilibrium cannot exceed a simple bound \cite{Warren1997,Jones2001} of
\begin{equation}\label{eq:warrenpop}
\frac{2\sinh(nh\nu/2kT)}{2^n\cosh^n(h\nu/2kT)}\approx\frac{n}{2^n}\times\frac{h\nu}{kT}
\end{equation}
where the approximation applies in the \textit{high temperature limit}, that is $h\nu\ll{k}T$, which is appropriate in conventional NMR experiments.  The key part of this result is the first term of the high temperature result, which shows that the observable signal decreases exponentially with the size of the spin system \cite{Warren1997}.  This effect appears to limit NMR quantum computing based on pseudo-pure states to around 10--20 qubits \cite{Gershenfeld1997,Gershenfeld1997a}.

This practical concern about the scalability of NMR quantum computers is bad enough \cite{Jones2000a}, but has also led to more fundamental concerns \cite{Braunstein1999} as to whether NMR quantum computers are really quantum devices at all!  When considering this claim it is vital to understand that the word \textit{quantum} is being used in a very technical sense, essentially equivalent to \textit{provably non-classical}, to describe systems which transcend known limits on the processing power of classical systems, and so the two formulations commonly heard (NMR quantum computers are not really quantum, and NMR quantum computers are quantum, but not usefully so) are less different than it might at first appear.

The essence of this discussion revolves around the problem of quantifying the extent of entanglement in pseudo-pure states.  Entanglement is widely believed to play a key role in quantum information processing \cite{Ekert1998}, and it has been established that many quantum algorithms must involve the generation of entangled states \cite{Linden2001}.  Unfortunately pseudo-pure states corresponding to entangled states (sometimes called \textit{pseudo-entangled states}) are only genuinely entangled if their purity is high enough \cite{Braunstein1999}.  It should be noted, however, that although individual NMR spin states can be described using a classical model, attempts to describe complete NMR experiments in this way \cite{Schack1999} have so far proved unsuccessful.

Problems of this kind were first considered by Werner \cite{Werner1989}, and two-qubit pseudo-pure states where the pure component is a Bell state are usually known as Werner states.  These states were analysed in detail by Peres \cite{Peres1996} who showed that a pseudo-entangled state of this kind is entangled if $\epsilon>\frac{1}{3}$ and is separable at or below this bound.  (Note that the bound corresponds to a fractional population of one-half in the Bell state, not one-third, as the two-thirds part of the ensemble in the maximally mixed state can itself be decomposed as an equal mixture of the four Bell states.)

Quantifying the entanglement of pseudo-pure states in larger spin systems is not a completely solved problem, but several key results have been derived.  Braunstein and coworkers \cite{Braunstein1999} originally derived a lower bound
\begin{equation}\label{eq:epslow}
\epsilon\le\frac{1}{1+2^{2n-1}}\sim\frac{2}{4^n}
\end{equation}
below which all pseudo-pure states are certainly separable, and an upper bound
\begin{equation}\label{eq:epshigh}
\epsilon>\frac{1}{1+2^{n/2}}\sim\frac{2}{2^{n/2}}
\end{equation}
above which provably entangled states can be made, leaving an intermediate region which is not well understood.  Subsequent results have tightened these bounds further \cite{Gurvits2003,Aubrun2006,Hildebrand2007}, but they do not as yet coincide.

\subsection{Preparing pseudo-pure states}
Although the basic idea of using pseudo-pure states is straightforward, it is necessary to consider how they can be prepared.  For a single spin the thermal state, with an excess population in the lower level, \textit{is} a pseudo-pure state, but this is not true in larger spin systems \cite{Jones1998b}.  Furthermore, in such systems the thermal state cannot be converted into a pseudo-pure state by any sequence of pulses and delays, or indeed by any unitary transformation.  This is most easily seen by noting that the eigenvalues of the two states are quite different, and that the eigenvalues of a matrix are invariant under unitary transformations.  (For a pseudo-pure ground state these eigenvalues are simply the diagonal elements of the density matrix, that is the populations of the basis states, while a general pseudo-pure state will also be diagonal in some appropriately chosen basis.)

A pseudo-pure state has a single large eigenvalue, $\epsilon+(1-\epsilon)/2^n$, corresponding to the population of the desired ground state, and $2^n-1$ identical smaller eigenvalues, $(1-\epsilon)/2^n$, corresponding to the equal populations of the various excited states; by contrast a thermal state has a complex pattern of eigenvalues reflecting the varying populations of the different excited states.  For example, for a homonuclear two-spin system the relative thermal state populations in the deviation density matrix are
\begin{equation}
\{1,0,0,-1\}
\end{equation}
while the desired pattern for a pseudo-pure state is
\begin{equation}
\{1,-\third,-\third,-\third\}.\label{eq:pptwo}
\end{equation}
This comparison also makes clear how in principle a pseudo-pure state can be prepared: all that is necessary is to average the values of the smaller eigenvalues while leaving the largest eigenvalue untouched.

\subsection{Temporal averaging}
Perhaps the conceptually simplest approach to averaging the smaller eigenvalues is the method of temporal averaging \cite{Knill1998}, in which the final result is obtained by averaging the experiment over a number of different starting states.  As quantum logic gates and the NMR detection process are linear in the input state, this is entirely equivalent to performing the experiment on a single averaged input state.  By this means the desired component can be retained, while the undesirable terms cancel out.  Conceptually this is somewhat similar to the use of phase cycling, with the averaging occuring over a range of different initial states.

For example, in a two-spin system, a pseudo-pure state can be achieved by averaging over the three initial states (described as before by the relative populations in their deviation density matrices)
\begin{equation}
\{1,0,0,-1\}\quad\{1,0,-1,0\}\quad\{1,-1,0,0\}
\end{equation}
as these average to the desired pseudo-pure state, Eq.~\ref{eq:pptwo}.  This example also makes clear which initial states should be averaged: the first state is simply the thermal state, while the remaining two can be obtained by cyclically permuting the three smaller populations in the thermal state.

In a system of $n$ spins there will be $2^n-1$ smaller populations in the thermal state $\rho_\textrm{th}$, and it clearly suffices to average the $2^n-1$ cyclic permutations of these populations to obtain the desired pseudo-pure state $\rho_\textbf{0}$. In general one may write
\begin{equation}
\rho_\textbf{0}=\frac{1}{2^n-1}\sum_{j=0}^{2^n-2}P_j\rho_\textrm{th}P^\dag_j
\end{equation}
where the $P_j$ are permutation operations, with $P_0$ being the trivial identity operation.  This approach is called exhaustive averaging, and will work for any size of spin system and for any set of thermal populations (so it will work for heteronuclear spin systems as well as homonuclear systems), or indeed for almost any set of initial populations.  All that is necessary is that the desired state $\ket{\textbf{0}}\bra{\textbf{0}}$ has the largest population in the system; if this is not the case then each permutation must be preceded by an initial operation which swaps the largest population to $\ket{\textbf{0}}\bra{\textbf{0}}$.  The desired permutation operations can be implemented using gate networks \cite{Knill1998,Kawamura2004}.  For the two-spin case the two permutation networks can be constructed using \CNOT\ gates
\begin{equation}
\raisebox{-2ex}{$P_1=$}\mbox{
\Qcircuit @C=1em @R=.7em {
& \ctrl{1} & \targ & \qw \\
& \targ & \ctrl{-1} & \qw
}}\qquad
\raisebox{-2ex}{$P_2=$}\mbox{
\Qcircuit @C=1em @R=.7em {
& \targ & \ctrl{1} & \qw \\
& \ctrl{-1} & \targ& \qw
}}
\end{equation}
and general permutations can always be constructed from \NOT, \CNOT\ and \TOFFOLI\ gates, although this theoretical description assumes that permutations can be implemented without errors arising from experimental imperfections and relaxation, which may not be realistic in complicated spin systems.

An obvious disadvantage of exhaustive temporal averaging is that the number of initial states used increases exponentially with the number of qubits in the system.  This increases the number of separate experiments by the same margin, and thus the overall time taken to implement the calculation; this completely cancels any exponential improvements in computational complexity that can arise from using quantum algorithms.  Because of this limitation there has been interest in finding more efficient methods, using non-cyclic permutations and applying unequal weightings to different permutations.  For example, preparing a pseudo-pure state in a system of four spins would require 15 cyclic permutations, but a method for achieving this with a weighted sum of only five permutations has been described \cite{Mori2005}.  For very large spin systems randomly chosen permutations can be used to prepare approximations to pseudo-pure states \cite{Knill1998}.

An alternative approach to temporal averaging is to decompose the desired pseudo-pure state in terms to product operators.  For example the deviation density matrix for the pseudo-pure ground state of a two-spin system (Eq.~\ref{eq:pptwopo}) contains the three product operators $I_z$, $S_z$ and $2I_zS_z$.  Each of these can be prepared as the starting state for an individual experiment, and the final result obtained by averaging over these starting states as before.  Clearly there are $2^n-1$  product operators needed to assemble a pseudo-pure state in an $n$ spin system (the spin system is described by $2^n$ product operators, but $\half{E}$ is not required), suggesting that $2^n-1$ separate experiments will be required, although in practice this can be reduced by combining experiments (obviously $I_z$ and $S_z$ can be ``prepared'' simultaneously).  In some cases it can be shown by pre-calculation that it is not necessary to include the contributions from certain initial states, as these cannot lead to any detectable NMR signals \cite{Marx2000}, but clearly this approach cannot be used in general.

\subsection{Spatial averaging}
Spatial averaging preceded temporal averaging as a method for preparing pseudo-pure states, but has generally proved less popular, perhaps because it requires magnetic field gradients which are not always available.  Conceptually it can be related to temporal averaging much as gradient coherence pathway selection is related to phase cycling, although this similarity is not always obvious in particular implementations.  The original approach of Cory \textit{et al.} \cite{Cory1996,Cory1997} uses \textit{crush gradients} to remove undesirable off-diagonal terms, leaving the desired diagonal density matrix.  It is most easily understood using product operators
\begin{equation}
\begin{split}
I_z+S_z
\xrightarrow{\makebox[2.5em]{$\scriptstyle60^{\circ}S_x$}}
&I_z+{\smhalf}S_z-\raisebox{0.4ex}{$\scriptstyle{\frac{\sqrt3}{2}}$}S_y\\
\xrightarrow{\makebox[2.5em]{$\scriptstyle\text{crush}$}}
&I_z+{\smhalf}S_z\\
\xrightarrow{\makebox[2.5em]{$\scriptstyle45^{\circ}I_{x}$}}
&\raisebox{0.4ex}{$\scriptstyle{\frac{1}{\sqrt2}}$}I_z
-\raisebox{0.4ex}{$\scriptstyle{\frac{1}{\sqrt2}}$}I_y+{\smhalf}S_z\\
\xrightarrow{\makebox[2.5em]{$\scriptstyle\text{couple}$}}
&\raisebox{0.4ex}{$\scriptstyle{\frac{1}{\sqrt2}}$}I_z
+\raisebox{0.4ex}{$\scriptstyle{\frac{1}{\sqrt2}}$}2I_xS_z+{\smhalf}S_z\\
\xrightarrow{\makebox[2.5em]{$\scriptstyle45^{\circ}I_{-y}$}}
&{\smhalf}I_z-{\smhalf}I_x+{\smhalf}2I_xS_z+{\smhalf}S_z+{\smhalf}2I_zS_z\\
\xrightarrow{\makebox[2.5em]{$\scriptstyle\text{crush}$}}
&{\smhalf}I_z+{\smhalf}S_z+{\smhalf}2I_zS_z
\end{split}
\end{equation}
where ``couple'' indicates evolution under the spin--spin coupling Hamiltonian $\pi{J}\,2I_zS_z$ for a time $1/2J$.  The coupling period can be implemented by refocusing the Zeeman interactions with spin echoes, but it is not in fact necessary to do so, as the initial state commutes with $S_z$ and evolution under the $I_z$ term can simply be tracked and then implemented by adjusting the phase of the subsequent pulse.  In general the action of the gradient pulses in crushing off-diagonal terms (which are sensitive to the phase of the reference frame) can significantly simplify the implementation of the transformations needed to prepare desired states.

As with any use of gradient crush sequences it is necessary to guard against inadvertent gradient echoes, where two crush sequences partly cancel each other causing undesired terms to be refocused.  In homonuclear systems it is also important to ensure that the off-diagonal terms do not include zero-quantum coherences, as these are not removed by crush gradients.  In heteronuclear systems this is not a problem, and the simpler sequence \cite{Pravia1999}
\begin{equation}
\begin{split}
I_z+S_z &\xrightarrow{45^{\circ}(I_x+S_x)}
\xrightarrow{\text{couple}}
\xrightarrow{30^{\circ}(I_{-y}+S_{-y})}\\
&\xrightarrow{\text{crush}}
\sqrt{\frac{3}{8}}\left(I_z+S_z+2I_zS_z\right)
\end{split}
\end{equation}
can be used instead.  It is, however, necessary to begin by equalising the polarisations of the I and S spins; this can be achieved either by applying a pulse with an appropriately chosen flip angle to the spin with higher polarization followed by a crush gradient, or by using a more complex sequence \cite{Pravia1999} which averages the two polarizations.  Sequences have also been developed for systems with larger numbers of spins \cite{Sakaguchi2000}, but these have not been widely used.

Spatial averaging is in some ways preferable to temporal averaging in that the result is obtained in a single experiment, but suffers from the obvious disadvantage that the signal intensity will be lower.  This is not only because temporal averaging combines data from multiple experiments, but also because the permutation method automatically produces the pseudo-pure state with the highest possible purity, while spatial averaging methods frequently sacrifice some of the available purity for simplicity in the preparation sequence.  One exception to this is use of controlled-transfer gates \cite{Kawamura2010}, which permit the spatial averaging equivalent of permutations; related ideas have been explored using line-selective pulses \cite{Peng2001}.  It is also possible to combine ideas from temporal and spatial averaging \cite{Yang2002}.

\subsection{Logical labelling}
Logical labelling \cite{Gershenfeld1997,Chuang1998b} is the conceptually most elegant approach to generating pseudo-pure states, although it is quite complex in practice and has not yet found widespread use.  The basic idea is to find a subset of spin states with a pattern of populations matching that of a pseudo-pure state, and then simply ``relabel'' these.  For example, a pseudo-pure state of a two qubit system has one state with a large population and three other states with equal smaller populations, and these can be mapped onto the ground state \ket{000} and the three excited states \ket{001}, \ket{010} and \ket{100} of a homonuclear three-spin system.  More sensibly one should choose the ground state \ket{000} and the three upper excited states \ket{110}, \ket{101} and \ket{011}, as this gives a higher signal intensity in the pseudo-pure state.  Clearly one must always embed the logical spin system within a larger number of physical spins, but the required overhead is not large \cite{Gershenfeld1997}.

As described the preparation sequence is extremely simple, comprising nothing more than mental relabelling of states, but this apparently simple approach would come at a great cost in the complexity of implementing quantum logic gates, as gates which act in a simple way on the logical qubits will be very complex when applied to the physical qubits.  A far simpler and more sensible approach is to permute the initial populations, so that the desired population pattern is moved to the four states \ket{000}, \ket{001}, \ket{010} and \ket{011}, giving a direct relationship between logical and physical states \cite{Gershenfeld1997,Chuang1998b}.  In this case the spin-system is in a pseudo-pure state, conditional on the first qubit being in state \ket{0}, and logic gates can now be implemented directly as long as they do not interchange states in the \ket{0} and \ket{1} manifolds of this labelling spin.  Explicit methods for doing this have been described \cite{Chuang1998b}.  A key disadvantage of logical labelling remains, however, in that it is necessary to ``waste'' at least one spin as a labelling spin.  Applications to date have concentrated on encoding two logical qubits in a three-spin system \cite{Dorai2000a,Vandersypen1999,Dorai2000}.

Finally there has been interest in combining logical labelling with averaging techniques, allowing the simple preparation of a pseudo-pure state of $n-1$ qubits in an $n$ spin system.  The labelled temporal averaging process is usually described as labelled flip and swap \cite{Knill1998}, and the spatially averaged logical labelling technique (SALLT) is broadly similar \cite{Mahesh2001a}.

\subsection{Other routes to pseudo-pure states}
In addition to these three main approaches, many other techniques have been described for preparing pseudo-pure states, or approximations to them.  Although superficially different from one another, many of these can be related to each other and to the approaches described previously \cite{Sharf2000}.  One interesting approach is based on selecting highly entangled states  \cite{Knill2000}, sometimes known as \textit{cat states} after Schr\"odinger's cat.  Such states can be easily generated from pure states using simple networks, as illustrated
\begin{equation}
\mbox{
\Qcircuit @C=1em @R=.7em {
\lstick{\ket{0}}&\gate{\textrm{H}}& \ctrl{1} & \qw& \qw&\qw \\
\lstick{\ket{0}}&\qw& \targ & \ctrl{1} & \qw&\qw&\rstick{(\ket{0000}+\ket{1111})/\sqrt{2}}\\
\lstick{\ket{0}}&\qw&\qw&\targ&\ctrl{1}&\qw\\
\lstick{\ket{0}}&\qw&\qw&\qw&\targ&\qw\gategroup{1}{6}{4}{6}{1em}{\}}
}}
\end{equation}
for a system with four qubits.  Such cat states are closely related to, although not quite the same as \cite{Jones1998b,Jones2001a}, maximal quantum coherences in conventional NMR; in particular they have the same $n$-fold sensitivity to frequency offsets and to phase shifts, and so gradients or phase cycling can be used to select them from complicated mixtures of states.  (This $n$-fold sensitivity of cat states is also the basis of entanglement assisted magnetic field sensing \cite{Jones2009,Simmons2010}.)

If the cat state preparation sequence is applied to a highly mixed state, such as a thermal equilibrium state, then the component in the desired ground state will be converted to a cat state, while other components will be converted to states with different coherence-orders.  Selecting for the cat state is then equivalent to selecting the ground state component, and reversing the cat state network will generate the desired ground state; the remaining components are converted into maximally mixed states, and so this process prepares a pseudo-pure ground state.

Unfortunately it is not possible to apply this procedure exactly as described: although cat states do uniquely show an $n$-fold sensitivity to phase shifts, the sensitivity is the same for the two cat states $(\ket{\textbf{0}}\pm\ket{\textbf{1}})/\sqrt{2}$.  Selecting for both these states is equivalent to selecting both the desired ground state and a second population state, corresponding to a pseudo-pure state of $n-1$ spins in an $n$-spin system \cite{Knill2000} with the remaining spin being in a state proportional to $\sigma_z$.  In some cases this apparent waste of a single spin is not, in fact, problematic \cite{Cummins2002} as the mixed state spin can still be used.  Despite this disadvantage, generation of maximum quantum coherences provides a natural route to the preparation of pseudo-pure states, and this approach has been explored in the context of dipolar coupled systems \cite{Lee2004,Furman2006}.  It is in general much simpler to produce states which are almost pseudo-pure than full pseudo-pure states and several techniques for doing this have been described \cite{Fung2001a,Fung2004,Peng2004}.  These ideas have also been explored in liquid crystal systems \cite{Lee2004,Lee2005,Lee2005a,Lee2007}.

Instead of preparing pseudo-pure states, it is possible to select a pseudo-pure component during the final detection of the NMR signal.  In some cases the signal from the desired pseudo-pure state is localised to a single line in a multiplet \cite{Jones2009,Simmons2010}, making its selection particularly simple.  A more complex approach is to use state tomography, described in Section~\ref{sec:tomography}, to completely characterise a mixed state, and then use an eigenvector analysis to select the component corresponding to the ground state \cite{Laflamme1998}.

\subsection{Non-Boltzmann states}
The approaches described above are all methods for preparing pseudo-pure states from a thermal mixed state in the high temperature limit ($kT\gg{h}\nu$), as this is only regime accessible at equilibrium in the liquid state.  To move beyond this it is necessary to prepare non-equilibrium states, in which the spin temperature is very different from the bulk temperature.  Here I will consider techniques for generating low temperature spin states, so that $kT<h\nu$.  Note that unless the spin temperature is very low, it will normally still be necessary to prepare a pseudo-pure state, but high polarization pseudo-pure states can in principle sustain entanglement and so such methods can be scalable.

Although there are a wide range of conventional NMR techniques for enhancing spin polarisation \cite{Jones2000a,Overhauser1953,Sorensen1989,Hore1993,Frydman2007}, very few of these are capable of reaching the low temperature regime in the liquid state.  The two main candidates for doing this are \textit{optical pumping} and the use of \textit{para-hydrogen}; a third possibility is the use of \textit{algorithmic cooling} to cool a small subset of spins in a larger spin system.

\subsection{Optical pumping}
Spin-exchange optical pumping of noble gas nuclei \cite{Walker1997}, notably \nuc{3}{He}, \nuc{129}{Xe} and \nuc{131}{Xe} has been used in a variety of NMR experiments, such as NMR imaging \cite{Albert1994}.  The process involves optical pumping of the electronic states of alkali metal atoms, followed by spin exchange between the electron and nuclear spin states. Noble gases are, however, unsuitable for NMR quantum computers as they are isolated atoms, and so can only provide single qubit systems. It is in principle possible to transfer the polarization to more complex molecules by cross polarization \cite{Navon1996}, but although this approach has been used to enhance signal intensities in NMR quantum computing \cite{Verhulst2001} the transfer efficiency is too low to be genuinely useful.  Optical pumping in bulk semiconductors \cite{Lampel1968} is even less likely to be useful for liquid state NMR computers.

\subsection{Para-hydrogen}
Para-Hydrogen Induced Polarization (PHIP) \cite{Bowers1986,Natterer1997,Duckett1999} is the only technique which has so far been used to prepare NMR spin systems with purities above the entanglement threshold \cite{Anwar2004}.  The properties of para-hydrogen are a consequence of the Pauli principle, which requires that $\textrm{H}_2$ molecules in  even rotational states have an antisymmetric nuclear wave function, that is the unique nuclear spin singlet state, while molecules in odd rotational states have nuclear spin triplet states.  Preparation of pure para-hydrogen is difficult because interconversion of singlet and triplet states is forbidden; however adsorption onto a catalyst surface breaks the symmetry, allowing interconversion to occur. Upon moving away from the surface interconversion is once again suppressed, and so the molecular spin state effectively remembers the temperature of the last catalytic surface encountered, and remains stable at room temperature.  A preparation temperature around 20\,K is low enough to prepare hydrogen with an essentially pure nuclear spin state \cite{Anwar2004}.

The para-hydrogen molecule cannot be observed directly by NMR because of its high symmetry, but it can be made to undergo a chemical reaction, producing a new molecule in which the two hydrogen atoms are distinct and can be observed and addressed.  In conventional PHIP experiments this reaction is fairly slow, taking around 1\,s, and the off-diagonal terms in the initial density matrix dephase at a rate determined by the frequency difference between the two spins, thus converting a pure nuclear spin singlet state into a mixture of singlet and $\textrm{T}_0$ triplet states.  In order to prepare a pure ground state for NMR quantum computing it is essential to suppress this dephasing.  One possibility is to apply an isotropic mixing sequence during the reaction \cite{Hubler2000}, but relaxation during the reaction remains a problem.

The most direct approach is to ensure that the reaction is rapid compared with dephasing and relaxation, and this can be achieved by using a rapid addition reaction with an unstable intermediate prepared by flash photolysis \cite{Anwar2004}.  This leads to an essentially pure spin singlet state which can easily be converted to a pure ground state \ket{00}, and used for quantum computation \cite{Anwar2004,Anwar2004b,Anwar2004a,Blazina2005}.  So far, however, the process has only be used to prepare two-qubit computers, although this polarization could in principle be shared among three qubits \cite{Anwar2006} while remaining above the entanglement threshold.

\subsection{Algorithmic cooling}
Algorithmic cooling refers to a family of techniques which use computational ideas to cool part of a spin system, transferring the undesired heat either to other parts of the spin system, or to the wider surroundings.  The original idea \cite{Schulman1999} is based on the observation that a molecular heat engine can be made to operate reversibly, and so can be implemented using only unitary quantum logic gates.  Similar ideas were previously considered in the context of polarisation transfer \cite{Sorensen1989}, but by using the full control offered by universal quantum logic gates it is possible to surpass the S{\o}rensen bounds, and generate spins with arbitrarily high polarisation.  The key step in this process, a \FREDKIN\ gate, or controlled-\SWAP, has been demonstrated in a three-spin system \cite{Chang2001}.

Although these methods do in principle allow arbitrarily high polarisations to be reached, the overhead needed is extraordinarily high \cite{Schulman1999}, with about $10^9$ thermal spins being needed to produce a single fully polarised spin \cite{Chang2001}.  Fortunately more practical approaches have been developed, with smaller overheads.  Algorithmic cooling \cite{Boykin2002} surpasses Shannon's spin entropy bound by permitting the hot spins to rethermalise with a lattice, so that the entropy is now conserved over the whole system and not just the spins in the quantum computer.  To work efficiently this process requires that the spins at the hot end of the molecular heat engine have much shorter $\textrm{T}_1$ times that the spins which are being cooled.  The basic ideas have been demonstrated in experiments \cite{Baugh2005,Ryan2008a}, and more sophisticated algorithms have been developed \cite{Fernandez2004,Schulman2005,Elias2006,Schulman2007,Elias2007}, but to date genuinely useful polarization enhancements remain out of reach.

\subsection{Use of mixed states}
The simplest approach to the problem of preparing pseudo-pure states is just to sidestep it completely by using mixed states directly.  One example is entanglement assisted magnetic field sensing, where it is not necessary to begin by preparing a pseudo-pure state as the desired component can be immediately identified as the signal from one of the outermost lines in a multiplet and the remaining lines can also be interpreted in a useful way \cite{Jones2009,Simmons2010}.

A more generally useful approach is to note that in many computations it is not necessary to prepare all the qubits in pseudo-pure states; instead some qubits can be left in mixed states \cite{Cappellaro2007a}.  Indeed it has been shown \cite{Knill1998a} that useful quantum computations can be performed in systems with a single pure qubit, with the other qubits being in maximally mixed states, an approach known as DQC1 (deterministic quantum computing with one quantum bit).  Although DQC1 is less powerful than a general purpose quantum computer, it can perform some tasks exponentially faster than the best known classical algorithm \cite{Poulin2004,Datta2005,Datta2007,Datta2008}.  If the DQC1 model is expanded to permit projective quantum measurements on a single qubit then it can even be used to implement Shor's quantum factoring algorithm \cite{Parker2000}; however this is less impressive than it might seem as the ability to perform projective measurements on a single qubit, when combined with quantum \SWAP\ operations, permits a pure state of the whole system to be prepared.

\section{Readout}
The final stage in a quantum computation is reading out the answer, which requires the final quantum state to be characterised.  This would normally be achieved by performing quantum measurements, which project the quantum state onto the measurement basis.  This approach is not possible in conventional NMR experiments, as detection of the NMR signal is not a strong projective measurement, but rather is a weak ensemble measurement \cite{Aharonov1990} which monitors the state of the spin system without changing it.  This is easily seen by noting that observing the free induction decay in an NMR experiment is equivalent to continuously monitoring two non-commuting observables, $\sigma_x$ and $\sigma_y$.  The explanation for this behaviour lies in properties of the operator corresponding to the mean of some observable evaluated over an ensemble of $N$ identical subsystem; in particular it can be shown that in the limit $N\rightarrow\infty$ every state is an eigenstate of every mean operator \cite{Aharonov1990}.  For finite values of $N$ the accuracy of this approximation grows with $\sqrt{N}$, and for NMR systems the error is negligible.

This ability to monitor without disturbance might seem to be an advantage of NMR over other, non-ensemble, approaches, but is in fact a source of many problems.  In particular, measuring a qubit is the best way of initialising it as discussed in Section~\ref{sec:initialisation}.  It is, however possible to demonstrate many quantum algorithms and phenomena using only weak measurements.  Just as for initialisation, a range of approaches have been used, but all these approaches combine the same main elements, based around the analysis of NMR spectra. For simplicity I will start by assuming that the computation ends with all the qubits in eigenstates, rather than in superposition states or entangled states; a small number of measurements will then provide all the information required.  A more thorough approach is to characterise completely the final state of the spin system by \textit{quantum state tomography}; while this can be useful in small spin systems the effort required for full tomography increases greatly with the size of the spin system

\subsection{Analysis of spectra}
The simplest situation is a one-qubit NMR quantum computer for which the states \ket{0} and \ket{1} correspond to the NMR states $I_z$ and $-I_z$ respectively. A $90^\circ\,I_y$ pulse will convert these to $\pm I_x$, which will appear as absorption or emission lines in a properly phased NMR spectrum; this makes sense as \ket{0} and \ket{1} correspond to excess population in the low energy and high energy spin states. As the absolute phase of an NMR signal is meaningless it is essential to obtain a reference signal against which phases can be determined; this is most easily achieved by acquiring a signal from the computer when it is known to be in the initial state \ket{0}.

The situation is similar with larger spin systems, but it is necessary to be more careful.  The NMR spin state corresponding to \ket{00} is not $I_z+S_z$ but $I_z+S_z+2I_zS_z$.  A general pseudo-pure eigenstate of two qubits can be expressed similarly as
\begin{equation}\label{eq:rhoab}
\ket{ab}\bra{ab}=\half\left((-1)^a I_z + (-1)^b S_z + (-1)^{a+b}\,2I_zS_z\right).
\end{equation}
This state can be analysed in two ways. In a homonuclear spin system it is simple to excite and observe both spins, and the two population terms ($I_z$ and $S_z$) are converted to observable single quantum coherences, while the longitudinal two-spin order ($2I_zS_z$) becomes unobservable double and zero quantum coherence.  The observable signal after a $90^\circ_y$ pulse is then
\begin{equation}
(-1)^a I_x + (-1)^b S_x
\end{equation}
and the desired state information is encoded in the phases (absorption or emission) of the NMR signals from the two spins.  In a heteronuclear spin system it is more natural to excite and observe just one spin, say $I$.  Application of a $90^\circ\, I_y$ pulse to the state gives
\begin{equation}
\left((-1)^a I_x + (-1)^b S_z + (-1)^{a+b}\, 2I_xS_z\right)/2,
\end{equation}
and the observable signal is now proportional to
\begin{equation}
(-1)^a \left( I_x + (-1)^b 2I_xS_z\right).
\end{equation}
Only one of the two lines in the $I$ spin doublet will be observed; the choice of line depends on $b$, the state of spin $S$, while the
phase of the signal indicates $a$, the state of spin $I$, as before.

\subsection{Forced decoherence} \label{sec:forceddecoherence}
The effect of a projective measurement on a qubit is to project a superposition of the form
\begin{equation}
\ket{\psi}=\alpha\ket{0}+\beta\ket{1}
\end{equation}
onto its eigenstates, so that qubit is found in  $\ket{0}$ with probability $|\alpha|^2$ and in $\ket{1}$ with probability $|\beta|^2$.  If the outcome of the measurement is unknown then the final state is described by a mixed state of the form
\begin{equation}
\rho=|\alpha|^2\ket{0}\bra{0}+|\beta|^2\ket{1}\bra{1}.
\end{equation}
Thus the key effect of projective measurement is to decohere the state, removing the off-diagonal coherence terms \cite{Zurek1991}.

The same result can be achieved without explicit measurements by forcibly increasing the decoherence rate.  This can be conveniently simulated by reducing the apparent coherence time $T_2^*$ by applying a magnetic field gradient.  Of course gradients do not truly decohere the state, as their effects
can be reversed by echoes, but diffusion within the sample volume means that gradients cannot be exactly reversed \cite{Sodickson1998}, and so gradient dephasing can be made indistinguishable from true decoherence \cite{Cory1998}.

Forced decoherence of this kind is widely used when it is desirable to simulate the effects of true projective measurements.  It can also be useful to remove undesired off-diagonal terms which arise from pulse sequence errors \cite{Jones1999a}, although in homonuclear systems it is necessary to worry about the invulnerability of zero-quantum error terms to gradient dephasing \cite{Jones1999a}.  This approach can be generalised to select other states; for example singlet states can be selected by \textsc{twirl} operations \cite{Anwar2005}.  Methods for simulating arbitrary decoherence processes have been described \cite{Havel2001}.

Artificial decoherence can also be imposed using temporal averaging rather than spatial averaging approaches, by averaging the result over experiments in which a range of gates are applied to the system \cite{Viola2001,Boulant2003}.  This approach allows considerable control over the exact form of the artificial decoherence, and also permits the decoherence to be applied rapidly within a particular instance of the experiment, although the use of temporal averaging means that the whole process takes longer than spatially averaged approaches.

\subsection{Tomography}\label{sec:tomography}
Quantum state tomography refers to the process of completely characterising the state of a quantum system.  Originally developed in optics \cite{Raymer1994}, the process is comparatively simple in spin systems where it is only necessary to determine a finite (if large) number of elements in the density matrix description.  Clearly this is not possible in a single experiment, and tomography can be quite a lengthy process.

In an NMR context it is most helpful to consider the process in terms of product operators \cite{Sorensen1983}.  A single spin is described by four operators, $\half{E}$, $I_x$, $I_y$ and $I_z$.  Two of these, $I_x$ and $I_y$, are NMR observables, and can be characterised directly by recording the free induction decay.  The $I_z$ term is not observable, but can be characterised by applying a $90^\circ$ pulse to convert it into one of the observable terms.  Note that this process inevitably renders one of the directly observable terms unobservable, and it is impossible to characterise these three terms in a single experiment.  The $\half{E}$ term is completely unobservable in any NMR experiment, and these methods only determine the deviation density matrix \cite{Chuang1998b}.  It is also necessary to obtain a reference spectrum from a spin in a known state (usually thermal equilibrium), firstly to obtain a phase reference allowing $I_x$ and $I_y$ to be separated, and secondly to obtain an amplitude reference so that the absolute size of the terms in the deviation density matrix can be determined.

In a two-spin system fifteen product operators have to be characterised \cite{Lee2002}, of which eight are directly observable, the four inphase coherences ($I_x$, $I_y$, $S_x$, $S_y$) and the four antiphase coherences ($2I_xS_z$, $2I_yS_z$, $2I_zS_x$, $2I_zS_y$).  In a homonuclear system these eight terms can all be determined from the amplitudes and phases of the four lines in the spectrum.  The remaining seven indirectly observable terms must be converted to observables, and it is not possible to achieve this for all seven terms in one go.  A minimum of four experiments is necessary \cite{Lee2002,Leskowitz2004}, although it is common to use a larger number.  One popular approach is to use nine experiments corresponding to three possible pulses (no pulse, $90^\circ_x$ and $90^\circ_y$) applied to spin $I$ and spin $S$ in all possible combinations \cite{Lee2002}.  Quantum state tomography has been widely demonstrated in spin-$\half$ NMR experiments \cite{Chuang1998b,Chuang1998,Roy2010,Yannoni1999,Das2003a,Mangold2004,Das2003}, as well as with quadrupolar nuclei \cite{Bonk2004,Bonk2005,Auccaise2008} and electron spins \cite{Scherer2008}.

It is possible to go beyond quantum state tomography to \textit{quantum process tomography} \cite{Chuang1997,Poyatos1997,Buzek1998}, which attempts to characterise the detailed behaviour of a pulse sequence by effectively determining the corresponding superoperator matrix.  This process is inevitably extremely time consuming, but has been demonstrated in liquid state \cite{Childs2001,Weinstein2004} and solid state \cite{Kampermann2005} NMR experiments.  Because of the extreme effort required for full process tomography there has been interest in partial tomography, which either seeks to determine the overall fidelity of the process \cite{Emerson2007,Levi2007,Knill2008} or concentrates on transfer between states of particular spins \cite{Cummins2002,Nielsen1998}.  As relaxation is a major source of errors in quantum processes there is also considerable overlap between quantum process tomography and studies of decoherence.

\subsection{Methods for single spin detection}
The prospects for NMR quantum computation would be greatly improved if it was in fact possible to perform conventional projective measurements on single spins.  Direct detection of NMR transitions of a single spin is impractical due to the very low energies involved, and it is necessary to couple the NMR transition to a process with a larger energy scale.  One possible approach is optically detected ENDOR \cite{Wrachtrup1997,Jelezko2002}, but most interest has focussed on Magnetic Resonance Force Microscopy (MRFM) \cite{Sidles1991,Rugar1992,Rugar2004}, and there have been proposals for building quantum computers with this approach  \cite{Berman2000a,Berman2001,Ladd2002}.  It should be noted, however, that such devices will necessarily bear little resemblance to current liquid state NMR experiments.  It has been suggested \cite{Jones2010} that spin-sensing chemical reactions can be considered as projective quantum measurements, but this interpretation is controversial.

A quite different perspective is provided by entanglement based spin measurements, which seek to enhance the signal from a single spin by correlating it with a large number of other spins \cite{Lee2005b}.  The basic method uses cat state techniques to entangle the single spin with a large number of \textit{ancilla} spins
\begin{equation}
\left(\alpha\ket{0}+\beta\ket{1}\right)\otimes\ket{0\dots0}\rightarrow\alpha\ket{00\dots0}+\beta\ket{11\dots1}
\end{equation}
and then performs a collective measurement on all the spins.  The cat-like state corresponds to a state of maximal quantum coherence, and is not directly detectable, but if the off-diagonal terms in the density matrix are dephased then the remaining diagonal terms correspond to population states with very high longitudinal spin order, which will give a correspondingly large collective signal.  This approach has been demonstrated in small systems \cite{Cappellaro2005} and possible implementations in much larger systems using techniques adapted from quantum cellular automata have been analysed \cite{Perez-Delgado2006,Kay2007,Lee2007a,Deng2008,Balachandran2009}.

\section{Decoherence}
Decoherence \cite{Zurek1991} refers to any incoherent processes which cause the spin system to undergo non-unitary evolution.  The term is essentially equivalent to relaxation.  Some authors chose to distinguish between $\textrm{T}_2$ processes, which they call decoherence, and $\textrm{T}_1$ processes, which they call relaxation, but such distinctions should not in general be relied upon.  As quantum computation relies on the use of coherent superpositions of quantum states, decoherence appears to provide a fundamental limit on the complexity of computations which can be performed \cite{Chuang1995}.  The development of \textit{decoherence free subspaces} \cite{Zanardi1997,Duan1997,Lidar1998,Lidar2001,Lidar2001a}, \textit{quantum error correction} \cite{Shor1995,Steane1996,Calderbank1996,Laflamme1996}, and \textit{fault tolerant computation} \cite{Knill1998b,Gottesman1998,Bacon2000} have enabled these limits to be partially sidestepped, but decoherence remains a fundamental concern.

\subsection{Modelling decoherence}
Although decoherence in large spin systems can be a very complex process, some progress can be made in simulations by incorporating very simple models of decoherence \cite{Vandersypen2001}, in essence assuming that every spin relaxes independently with its own time constants $\textrm{T}_1$ and $\textrm{T}_2$.  Typically relaxation is dominated by $\textrm{T}_2$ processes, but these largely lead to signal loss, without introducing significant error signals; by contrast $\textrm{T}_1$ processes, although usually significantly slower, can lead to ambiguous results in some quantum computations.

Both types of relaxation are easily modelled using the \textit{operator sum representation} \cite{NCbook}
\begin{equation}
\rho\rightarrow\sum_kE_k\rho E_k^\dag
\end{equation}
where the operators $E_k$ can take any form (in particular, they need not be unitary), subject only to the requirement that
\begin{equation}
\sum_kE_k^\dag E_k=\mathbf{1}
\end{equation}
if the process is to be trace preserving (so that density matrices remain density matrices).  $\textrm{T}_2$ processes are easily described using \textit{phase damping} \cite{Vandersypen2001}, which relies on the fact that for a single qubit the process
\begin{equation}
\rho\rightarrow\half\,\rho+\half\,Z\rho Z
\end{equation}
will completely remove any off-diagonal terms from a density matrix $\rho$.  More realistic decoherence can be modelled by removing the off-diagonal terms from some steadily increasing fraction of the density matrix.  This gives the set of operators
\begin{equation}
E_0=\sqrt{\lambda}\begin{pmatrix}1&0\\0&1\end{pmatrix}\qquad E_1=\sqrt{1-\lambda}\begin{pmatrix}1&0\\0&-1\end{pmatrix}
\end{equation}
where
\begin{equation}
\lambda=\half\left(1+\textrm{e}^{-t/T_2}\right).
\end{equation}
$\textrm{T}_1$ processes are a bit more complicated, as it is necessary to allow for the fact that the system does not relax to the ground state, but rather to the thermal equilibrium state.  This process, \textit{generalized amplitude damping} \cite{Vandersypen2001}, can be described by the set of four operators
\begin{equation}
\begin{split}
E_0=\sqrt{p}\begin{pmatrix}1&0\\0&\sqrt{1-\gamma}\end{pmatrix} &\quad E_1=\sqrt{p}\begin{pmatrix}0&\sqrt{\gamma}\\0&0\end{pmatrix}\\&\\
E_2=\sqrt{1-p}\begin{pmatrix}\sqrt{1-\gamma}&0\\0&1\end{pmatrix} &\quad E_3=\sqrt{1-p}\begin{pmatrix}0&0\\\sqrt{\gamma}&0\end{pmatrix}
\end{split}
\end{equation}
where
\begin{equation}
\gamma=1-\textrm{e}^{-t/T_1}\qquad p=\half(1+\epsilon)
\end{equation}
and $\epsilon$ is the polarization at thermal equilibrium, so that the thermal state is
\begin{equation}
\rho_\textrm{th}=\half{E}+\epsilon{I_z}.
\end{equation}
As expected $\rho_\textrm{th}$ is unaffected by either kind of relaxation.  It might seem tempting to concentrate on the pseudo-pure component of the spin state, which corresponds to $\epsilon=1$, in which case the operators $E_2$ and $E_3$ can be dropped, but while this process will work for a single spin it will give incorrect results when extended to larger spin systems.

The complete relaxation process for a single spin can be described by applying both amplitude damping and phase damping, remembering that the measured $\textrm{T}_2$ time includes a contribution from $\textrm{T}_1$ relaxation.  The relaxation of a group of two or more spins can be described by applying relaxation sequentially to each spin; in this case it is vital to calculate the amplitude damping correctly, so that the system relaxes back to the correct thermal state and not to a pseudo-pure state.

The approach outlined above only applies to decoherence during a single period of free precession.  With a more complicated pulse sequence the simplest approach is to break it up into a sequence of pulses (assumed to be instantaneous) and delays, with the decoherence that occurs during a delay being calculated at the end of the delay, immediately before the following pulse.  Note that only the spins experiencing the pulse need be considered at this stage; decoherence of other spins can be saved up, to be applied just before the next pulse addressing those spins.  If only $\textrm{T}_2$ processes are considered then this can be simplified further, as $\textrm{T}_2$ relaxation is not affected by $180^\circ$ pulses, and so $\textrm{T}_2$ relaxation can be calculated at the end of a whole spin-echo period, rather than step by step. This approach will not be entirely correct in experiments with long selective pulses, but has proved fairly successful in some cases \cite{Vandersypen2001}.  It will, however, be difficult to apply to pulse sequences which make extensive use of gates derived by optimal control methods, as in this case any division into pulses and delays is to some extent artificial.

\subsection{Decoherence in large spin systems} \label{sec:DLS}
The previous section has assumed that every spin relaxes individually, uncorrelated with the relaxation of neighbouring spins, and it is well known that this assumption is rarely correct in NMR spin systems.  Obvious examples in conventional NMR include nuclear Overhauser effects \cite{Noggle1971} arising from correlated $\textrm{T}_1$ relaxation, the differential linewidths in multiplets used in transverse relaxation-optimized spectroscopy (TROSY) experiments \cite{Pervushin1997}, and the relaxation of multiple quantum coherences \cite{Wokaun1978}.

%\subsection{Studying decoherence}
NMR techniques can also be used to study decoherence processes, and in particular to determine how decoherence rates scale with the number of qubits in real physical systems, as well as the use of decoupling techniques to reduce decoherence rates by suppressing undesirable interactions in such systems \cite{Kawamura2005,Kawamura2006,Krojanski2006,Tei2003,Krojanski2004,Lovric2007}.

%\section{Electron spins and ENDOR}
%\lid

\section{Quantum Algorithms}
I will now consider some of the quantum algorithms which have been implemented using NMR quantum computers.  Before doing so it is useful to say a little more about the implementation of mathematical operations with reversible quantum devices, and in particular to explore how functions can be evaluated in a reversible context.  The obvious direct approach of creating a propagator which implements
\begin{equation}
\ket{x}\overset{U_f}\longrightarrow\ket{f(x)}.
\end{equation}
is not normally practical, as this process will only be reversible if the function is itself intrinsically reversible, which is not generally the case.  The solution to this problem is to preserve the input qubits and store the output in an ancilla bit.  The output cannot, however, simply overwrite the initial value of the ancilla, as that would be irreversible. Instead the new value of the ancilla must be obtained by reversibly combining the output of the gate with the old value, and this is most simply achieved by using bitwise addition modulo 2 (the \XOR\ gate), so that
\begin{equation}
\ket{x}\ket{y}\overset{U_f}{\longrightarrow}\ket{x}\ket{y\oplus{f(x)}}.
\end{equation}
If the ancilla bit is initially set to zero then the value of the function can be easily obtained
\begin{equation}
\ket{x}\ket{0}\overset{U_f}\longrightarrow\ket{x}\ket{f(x)}.
\end{equation}
For a function with two input bits and one output bit this can be written as a quantum network
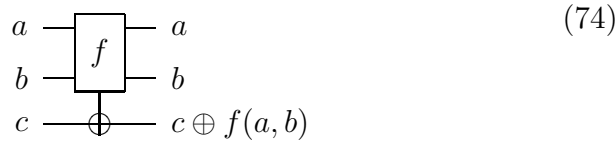
\begin{equation}
\mbox{ \Qcircuit @C=1em @R=0.7em {
\lstick{a} &\multigate{1}{f}&\rstick{a}\qw\\
\lstick{b} &\ghost{f}       &\rstick{b}\qw\\
\lstick{c} &\targ \qwx      &\rstick{c\oplus f(a,b)}\qw
} }
\end{equation}
sometimes called a $f$-controlled-\NOT\ gate. Functions with more than one output bit can be handled by combining each output bit with its own ancilla.  Note that while reversible logic can perform any desired transformation on a set of input bits it does not provide any means to set the input or ancilla bits into the desired initial states, and so a reversible computation must begin with an irreversible initialisation process.

\subsection{Deutsch's algorithms and related methods}
Deutsch's algorithm plays a central role in the historical development of quantum computing, being both the first quantum algorithm to be described and the first quantum algorithm to be implemented experimentally.  It considers the analysis of the simplest type of functions: binary functions which map a single input bit to a single output bit.  Clearly there are only four such functions
\begin{equation}
\begin{tabular}{|l|l|l|l|l|}
\hline $x$&$f_{00}(x)$&$f_{01}(x)$&$f_{10}(x)$&$f_{11}(x)$
\\\hline 0& 0 & 0 & 1 & 1\\ 1& 0 & 1 & 0 & 1\\\hline
\end{tabular}
\end{equation}
where each function is conveniently labeled by the pattern of output bits in its truth table.

Deutsch's problem considers the identification of a function, which is known to be one of these four functions but otherwise undetermined, in a situation where the only possible access to the function is by finding its output for a particular input (that is, there is no way of determining \textit{how} the function values are calculated).  This is sometimes called the \textit{oracle} model of computation, and each function evaluation is called an \textit{oracle query}.  Clearly it is necessary and sufficient to make two oracle queries to completely identify a function, evaluating both $f(0)$ and $f(1)$ with $f$-controlled-\NOT\ gates.

The function can in some cases be partially identified in a single query: for example $f_{00}$ and $f_{01}$ can be distinguished from $f_{10}$ and $f_{11}$ by evaluating $f(0)$.  Suppose, however, it is desired to distinguish functions according to their \textit{parity}, $f(0)\oplus{f}(1)$, so that the \textit{balanced} functions $f_{01}$ and $f_{10}$, which give outputs of 0 for one input and 1 for the other, are distinguished from the \textit{constant} functions $f_{00}$ and $f_{11}$, which give the same output for all inputs.  With a classical computer this distinction requires two queries, as calculating the parity requires both $f(0)$ and $f(1)$ to be determined.  Deutsch, however, showed that a quantum computer can determine the parity with a single query.

Deutsch's original algorithm is not particularly useful, as it only works half the time at random \cite{Deutsch1985}, reporting an inconclusive result the other half of the time.  Thus references to Deutsch's algorithm almost always refer to a more modern version which works with certainty \cite{Cleve1998}.  The quantum network to achieve this
\begin{equation}
\mbox{ \Qcircuit @C=1em @R=0.7em {
\lstick{\ket{0}} &\gate{\textrm{H}}&\gate{f}&\gate{\textrm{H}}&\rstick{\ket{f(0)\oplus{f(1)}}}\qw\\
\lstick{\ket{1}} &\gate{\textrm{H}}&\targ\qwx&\gate{\textrm{H}}&\rstick{\ket{1}}\qw
} }
\end{equation}
is well explained elsewhere \cite{Jones1998c,Jones2001a,Merminbook,Cleve1998}, but can be broken down into three parts.  The initial Hadamard gate on the first qubit generates $\ket{+}$, a uniform superposition of the two possible input values; this is then passed to the $f$-controlled-\NOT\, which evaluates the function for both inputs simultaneously.  As the ancilla qubit is in the state $\textrm{H}\ket{1}=\ket{-}$ the effect of the function on a single input is given by
\begin{equation}
\ket{x}\ket{-}=\frac{\ket{x}\ket{0}-\ket{x}\ket{1}}{\sqrt{2}}\overset{U_f}{\longrightarrow}\frac{\ket{x}\ket{f(x)}-\ket{x}\ket{1\oplus{f(x)}}}{\sqrt{2}}
\end{equation}
and this result can be simplified by considering the two possible values of $f(x)$.  For the case $f(x)=0$ the result is
\begin{equation}
\frac{\ket{x}\ket{0}-\ket{x}\ket{1}}{\sqrt{2}}=\ket{x}\ket{-}
\end{equation}
while if $f(x)=1$ the result is
\begin{equation}
\frac{\ket{x}\ket{1}-\ket{x}\ket{0}}{\sqrt{2}}=-\ket{x}\ket{-}.
\end{equation}
This can be summarised as
\begin{equation}
\ket{x}\ket{-}\overset{U_f}{\longrightarrow}(-1)^{f(x)}\ket{x}\ket{-}
\end{equation}
so the value of the function determines the phase of the state.  This might appear to be a global phase, and so irrelevant, but if the input qubit is in a superposition state then the function values can appear as a relative phase
\begin{equation}
\ket{+}\ket{-}\overset{U_f}{\longrightarrow}\frac{(-1)^{f(0)}\ket{0}+(-1)^{f(1)}\ket{1}}{\sqrt{2}}\ket{-}.
\end{equation}
If the function $f$ is constant, so that $f(0)=f(1)$, this simplifies to
\begin{equation}
(-1)^{f(0)}\frac{\ket{0}+\ket{1}}{\sqrt{2}}\ket{-}=(-1)^{f(0)}\ket{+}\ket{-}
\end{equation}
while if the function is balanced the result is
\begin{equation}
(-1)^{f(0)}\frac{\ket{0}-\ket{1}}{\sqrt{2}}\ket{-}=(-1)^{f(0)}\ket{-}\ket{-}.
\end{equation}
The effect of the final pair of Hadamard gates is to convert this to $\ket{0}\ket{1}$ or $\ket{1}\ket{1}$ as appropriate, completing the desired computation of the parity.  Note that the final value of $f(0)$ is encoded as an undetectable global phase, and so this algorithm enables \textit{only} the parity to be determined; it is not possible to make further distinctions between the two constant or two balanced functions without making further oracle queries.

The simplicity of the network, requiring only two qubits, makes it an extremely attractive early target.  It is necessary to implement the $f$-controlled-\NOT\ gates explicitly, but these are trivial for the two constant functions, $f_{00}$ and $f_{11}$, and only require a single \CNOT\ gate for the two balanced functions, $f_{01}$ and $f_{10}$ \cite{Jones1998c,Jones2001a}.  Deutsch's algorithm was implemented almost simultaneously using homonuclear \cite{Jones1998c} and heteronuclear \cite{Chuang1998} two-spin systems, and has subsequently been implemented on an ion trap quantum computer \cite{Gulde2003}, with a pure state NMR quantum computer \cite{Anwar2004b}, and with a photon cluster state quantum computer \cite{Tame2007}.

Most NMR developments have, however, involved not Deutsch's algorithm, but its generalisation, the Deutsch--Jozsa algorithm \cite{Cleve1998,Deutsch1992}, which considers functions from $n$ bits to 1 bit (as Deutsch's algorithm can be considered as the simplest example of the more general Deutsch--Jozsa algorithm, with $n=1$, the general name is sometimes used for this case as well \cite{Hubler2000,Zhang1999}).  Such functions need not be constant (the same output for all inputs) or balanced (an equal number of 0s and 1s in the outputs); for example a function with $n=2$ might output 0 for one input and 1 for three inputs.  Suppose, however, that the function is \textit{promised} to be either constant or balanced; in this case it is possible to determine which sort of function it is with a single evaluation.  The quantum network is an expansion of that for Deutsch's algorithm, and for $n=2$ takes the form
\begin{equation}
\mbox{ \Qcircuit @C=1em @R=0.7em {
\lstick{\ket{0}} &\gate{\textrm{H}}&\multigate{1}{f}&\gate{\textrm{H}}&\rstick{\ket{a}}\qw\\
\lstick{\ket{0}} &\gate{\textrm{H}}&\ghost{f}&\gate{\textrm{H}}&\rstick{\ket{b}}\qw\\
\lstick{\ket{1}} &\gate{\textrm{H}}&\targ\qwx&\gate{\textrm{H}}&\rstick{\ket{1}}\qw
} }
\end{equation}
where the output bits, $a$ and $b$, are both equal to 0 if the function is constant, and are not both equal to 0 if the function is balanced.  This generalises to $n$ bit functions in the obvious way, with all output bits equal to 0 if the function is constant, and at least one bit equal to 1 if the function is balanced.

Implementing the Deutsch--Jozsa algorithm would appear to require a device with at least three qubits, and the first implementation \cite{Linden1998} did precisely that, using line selective pulses in a three-spin homonuclear spin system; this simplifies the implementation of two of the balanced functions, which would require \TOFFOLI\ gates if implemented using standard circuits \cite{Dorai2000,Mahesh2001,Ermakov2003}.  The algorithm has also been implemented for 4-bit functions in a 5-spin system \cite{Marx2000}, restricting the choice to a single balanced function which could be implemented relatively simply.

Although the conventional Deutsch--Jozsa algorithm requires a minimum of three qubits, it was noted early on \cite{Collins1998} that the ancilla bit is not really necessary.  The ancilla bit is used in classical function evaluations to hold the function value, but in quantum versions the effect of setting the ancilla bit to \ket{-} is that the function value is stored as a phase rather than a bit value.  Thus it is possible to remove the ancilla qubit if the conventional classical oracle is replaced by a purely quantum oracle, which implements
\begin{equation}
\ket{x}\overset{U_f}{\longrightarrow}(-1)^{f(x)}\ket{x}
\end{equation}
directly.  This quantum oracle cannot be used for classical algorithms (as the qubit phase is a purely quantum mechanical phenomenon), but will work perfectly well in quantum algorithms.  This purely quantum version \cite{Dorai2001,Collins2000}, which is sometimes called the \textit{refined} Deutsch--Jozsa algorithm \cite{Kim2000} or the Collins version of the Deutsch--Jozsa algorithm \cite{Mangold2004}, permits $n$-bit functions to be studied using an $n$-spin system.  This approach can even be used to implement Deutsch's algorithm with a single spin \cite{Arvind2001}.

\subsection{Grover and other quantum search algorithms}
Grover's quantum search algorithm \cite{Grover1997,Jones1998a} is the simplest member of the second major family of quantum algorithms.  In general the algorithm concerns the analysis of functions of $n$ input bits for which the output is 0 except for a single \textit{satisfying} input, for which the output is 1.  If the function is only accessible through an oracle implementation, so that it is not possible to analyse the function itself, but only its values, the only practical approach to locating the satisfying input is to try inputs at random until it is found.  This unstructured search, equivalent to trying to find the name corresponding to a particular telephone number by reading through the telephone directory, is obviously extremely inefficient: for a function with $n$ input bits there are $N=2^n$ possible inputs, and on average it will be necessary to try about $N/2$ of these before the satisfying input is found.  By contrast, Grover's algorithm allows it to be located with around $\sqrt{N}$ queries.

The quantum circuit implementing Grover's search algorithm in the case $n=2$ ($N=4$) is
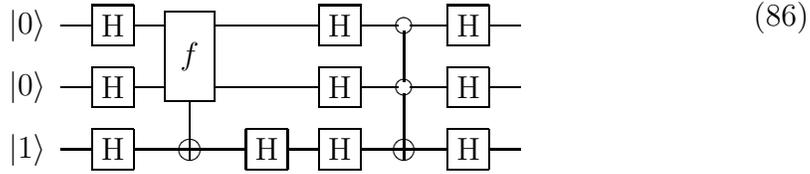
\begin{equation}
\mbox{ \Qcircuit @C=1em @R=0.7em {
\lstick{\ket{0}} &\gate{\text{H}}&\multigate{1}{f}&\qw            &\gate{\text{H}}&\ctrlo{1}&\gate{\text{H}}&\qw\\
\lstick{\ket{0}} &\gate{\text{H}}&\ghost{f}       &\qw            &\gate{\text{H}}&\ctrlo{1}&\gate{\text{H}}&\qw\\
\lstick{\ket{1}} &\gate{\text{H}}&\targ\qwx       &\gate{\text{H}}&\gate{\text{H}}&\targ    &\gate{\text{H}}&\qw
}}
\end{equation}
where the first two bits represent the inputs to the function and the third bit is an ancilla.  The first three qubit gate is an $f$-\CNOT\ gate and the second one is a close relative of the \TOFFOLI\ gate which applies a \NOT\ gate to the ancilla bit if the input bits are both in state 0.  As in the Deutsch--Jozsa algorithm the ancilla bit is not really necessary as it only serves to convert the function result from a bit value to a phase, and so the network can be simplified to
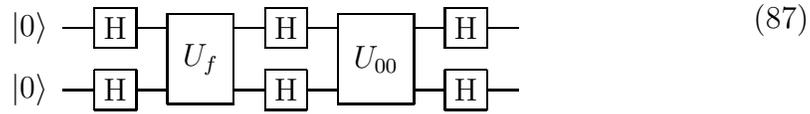
\begin{equation}
\mbox{ \Qcircuit @C=1em @R=0.7em {
\lstick{\ket{0}}&\gate{\text{H}}&\multigate{1}{U_f}&\gate{\text{H}}&\multigate{1}{U_{00}}&\gate{\text{H}}&\qw\\
\lstick{\ket{0}}&\gate{\text{H}}&\ghost{U_f}       &\gate{\text{H}}&\ghost{U_{00}}       &\gate{\text{H}}&\qw
}}
\end{equation}
where $U_{00}$ negates the basis state \ket{00} but leaves all other basis states alone, and $U_f$ negates the basis state corresponding to the unique satisfying input.  This simplified algorithm is universally used in NMR implementations of Grover's search.

As usual the initial Hadamard gates convert the input qubits to a uniform superposition of the four possible inputs, and the function is evaluated over all four inputs in parallel, leaving the satisfying input uniquely marked with a phase shift.  The remaining gates then convert this phase difference to an amplitude difference, so that all the amplitude in the superposition is concentrated on the satisfying input, and when the input bits are measured at the end of the process their values will indicate the satisfying input.  The circuit is similar for cases with $n>2$, except that it is now necessary to repeat the function evaluation and amplitude concentration stages several times, with the amplitude of the satisfying input in the final superposition varying sinusoidally with the number of repetitions.  This has two consequences: firstly that it is necessary to choose the number of repetitions appropriately, and secondly that it is usually not certain that the satisfying input will be located, even if the optimum number of repetitions are used, as in the general case the amplitude may never be completely concentrated on the satisfying input.

The case $n=2$ is special, in that the algorithm is guaranteed to work after a single repetition, making this case an attractive target, and Grover's search was implemented early on with both heteronuclear \cite{Chuang1998a} and homonuclear \cite{Jones1998d} spin systems.  Later implementations have been used to demonstrate the use of dipolar couplings in liquid crystal solvents \cite{Yannoni1999}, efficient quantum state tomography \cite{Das2003a}, the stability of Grover's search under imperfect gates \cite{Long2001a}, and NMR quantum computing with highly polarized \cite{Verhulst2001} and almost pure \cite{Anwar2004a} states.  Unlike the Deutsch--Jozsa algorithm there has been little interest in increasing the value of $n$, but there has been one demonstration of searching over three-bit functions on a three-spin device \cite{Vandersypen2000}, demonstrating the expected oscillatory behaviour as the number of repetitions is varied.

Grover's algorithm can be extended to more general functions with more than one satisfying input; the network is identical, but the number of repetitions required now depends on $\sqrt{N/k}$, where $k$ is the number of satisfying inputs \cite{Boyer1996}. The output qubits are in a uniform superposition of all the matching inputs, which makes this an unattractive algorithm for NMR implementations, as the non-projective readout process will not cope well with this superposition.  A related algorithm, approximate quantum counting, permits $k$ to be estimated by observing the oscillation frequency \cite{Boyer1996,Mosca2001}, and this is a much more attractive target.  The algorithm requires an additional control qubit in addition to the $n$ bits defining the input to the function, but a two-spin system has been used \cite{Jones1999a} to demonstrate the case $n=1$, for which $k$ equals 0, 1 or 2, and a related experiment has been performed in a three-spin system \cite{Lee2002a}.  The oscillatory behaviour inherent in quantum counting allows this algorithm to be used to explore the effects of applying very large numbers of logic gates, and this has been used to demonstrate the effects of composite pulses in suppressing off-resonance errors \cite{Cummins2000}.

An alternative approach developed by Grover replaces the simple repetition of a single search operator with a more complicated recursive operator \cite{Grover2005}, which causes the algorithm to converge on a superposition of satisfying inputs no matter what the value of $k$ is.  This algorithm permits an efficient search to be performed when the value of $k$ is unknown, although it is much less efficient than the original quantum search algorithm when $k$ is known.  This algorithm has been implemented by NMR \cite{Xiao2005} for the case $n=2$ in a two-spin system.  Results were demonstrated for functions with $k=1$ and $k=2$; in the latter case care was needed in interpreting the final result, as this corresponds to a superposition of the two satisfying states.

In addition to Grover's unstructured search, quantum algorithms have also been developed for searching functions whose structure is partly known.  NMR implementations include the Bernstein--Vazirani algorithm \cite{Das2004,Du2001} and Hogg's structured search \cite{Das2004,Zhu2001,Peng2002}.  Several of these implementations build on the concept of searching in Liouville space \cite{Madi1998,Bruschweiler2000,Long2003} which is particularly appropriate to ensemble implementations such as NMR.

\subsection{Shor type algorithms}
Shor's quantum factoring algorithm \cite{Shor1994,Ekert1996,Jozsa1998,Shor1999} is perhaps the best known application of quantum computation because of its potential significance to wider society.  Many current cryptographic schemes underlying the security of financial systems \cite{Graff2001} and of the internet in general \cite{Anderson2001} rely on the apparent difficulty of factoring large composite numbers, or closely related problems such as calculating discrete logarithms \cite{Schneier1995}.  No efficient algorithm is known for solving these problems on a classical computer, and it is widely presumed that no such algorithm exists.  By contrast a quantum computer can efficiently solve such problems, rendering much of current cryptography obsolete.

Shor's algorithm works by finding the period of the modular exponentiation function, after which results from classical number theory allow a variety of particular problems to be solved \cite{Ekert1996}.  Period finding is, in fact, a general feature of several quantum algorithms, such as the Deutsch--Jozsa algorithm, and Shor's algorithm can be related to these \cite{Jozsa1998}.  A key element is the implementation of a quantum Fourier transform \cite{Ekert1996} which was demonstrated by NMR in a three-spin system \cite{Weinstein2001} and used to find the order of a permutation in a five spin system \cite{Vandersypen2000a}.

It was then a small step to a full implementation of Shor's algorithm by NMR \cite{Vandersypen2001}.  A key step in achieving this \cite{Jones2002} was that realisation that the implementing the simplest non trivial example of the algorithm, in effect factoring $N=15$, is particularly simple, requiring only seven qubits.  The implementation was further simplified by noting that certain gates could be, in effect, compiled out of the quantum network, substantially reducing the complexity of the final circuit.  Even so the final pulse sequence \cite{Vandersypen2001}, containing around 300 distinct shaped pulses, remains one of the most complex NMR experiments ever demonstrated.

\subsection{Quantum simulation}
Quantum simulation is the idea of using one quantum mechanical system to simulate another quantum mechanical system more efficiently than can be achieved by any system based on classical physics \cite{Feynman1982}.  Although such simulations may not look like conventional algorithms, a quantum simulator can equally well be thought of as a special purpose quantum computer for solving an otherwise intractable problem.

It is wise to use some caution in developing this argument, as in principle almost any physical process can be considered as a computation \cite{Toffoli1982,Lloyd2006}.  The approach is, however, clearly useful as long as it is possible to simulate evolution under arbitrary Hamiltonians,  and it has been shown \cite{Lloyd1996,Abrams1997,Zalka1998,Lidar1999} that \textit{any} quantum system evolving under a local Hamiltonian can be efficiently simulated on a quantum computer.  Specific algorithms have been developed for many important problems, including molecular energies \cite{Aspuru-Guzik2005}, molecular geometries \cite{Kassal2009}, and chemical dynamics \cite{Kassal2008}.  Arbitrary Hamiltonians can either be simulated using quantum networks, or by constructing effective Hamiltonians using Trotter approximations \cite{Vandersypen2004,Trotter1958}.

Simple simulations can be implemented on a two qubit system, and NMR was used early on to simulate a truncated quantum harmonic oscillator \cite{Somaroo1999,Tseng2000}.  Subsequently NMR techniques were used to simulate the Heisenberg Hamiltonian in spin chains \cite{Peng2005}, and pairing Hamiltonians \cite{Wang2006,Yang2006a,Brown2006}.  Recently quantum chemistry algorithms developed for use in photonic systems \cite{Lanyon2010} have also been applied in NMR \cite{Du2010}.

\subsection{Adiabatic algorithms}
If a quantum system starts in the ground state of its Hamiltonian, and the Hamiltonian is slowly varied, then the system will stay in its instantaneous ground state.  This adiabatic following forms the basis of adiabatic algorithms for quantum computing \cite{Farhi2001,Childs2001a}.  These algorithms work by constructing a Hamiltonian corresponding to the problem to be solved, such that the ground state of the Hamiltonian encodes the solution.  This ground state can then be determined by slowly interpolating the experimental Hamiltonian between some initial form, such as the background Hamiltonian of the system, and the desired final form.  Although this approach appears quite different to conventional circuit methods, it has been shown that the two schemes are fundamentally equivalent \cite{Aharonov2007}, and the methods used to implement adiabatic algorithms are closely related to those used in quantum simulations.

This approach was first implemented using a three-spin system \cite{Steffen2003} to implement an optimization algorithm, with the interpolation between the two Hamiltonians implemented using Trotter approximations \cite{Trotter1958}.  Subsequently adiabatic methods were used to demonstrate Grover's quantum search and Deutsch's algorithm \cite{Mitra2005}, to solve the 1-SAT satisfiability problem \cite{Mitra2008}, and to implement quantum factoring \cite{Peng2008}.

\subsection{Gauss sums}
Several approaches have been described \cite{Mehring2007,Mahesh2007,Peng2008a} for implementing Gauss sums by NMR, and it has been suggested that this can be used as a method for factoring numbers.  This relies on the fact that the Gauss sum
\begin{equation}
\mathcal{A}_N^{M}(l) = \frac{1}{M+1} \sum_{m=0}^M \exp\left[- \mathrm{i}\,2\pi\, m^2\frac{N}{l}\right]
\label{eq:Gauss}
\end{equation}
(where $N$ is an integer to be factored, and $l$ is a trial factor)is equal to $1$ if $l$ is indeed a factor of $N$, and has an absolute value less than one otherwise, with the exact value depending on $M$. Several different methods for evaluating this sum have been described, but in essence they all involve applying a sequence of many pulses to a single isolated spin, with the phase of each pulse effectively encoding a single term from the Gauss sum. Related experiments have also been implemented with other techniques \cite{Gilowski2008,Bigourd2008}.

These NMR methods have generated considerable interest, as they can be applied to far larger numbers than current implementations of Shor's quantum factoring algorithm. It is, however, important to note that Gauss sums do not in themselves provide a factoring algorithm, but rather a factor checking algorithm, and it may be necessary to try all possible prime factors up to $\sqrt{N}$.  Since the implementations only involve a single spin they cannot, of course, involve entangled states, raising further doubts as to their importance.  However it has been previously noted that even single qubits using superpositions to integrate phase can outperform simple classical systems \cite{Galvao2003}. Unfortunately current NMR methods to evaluate Gauss sums require the initial computation of pulse phases, and this cannot be implemented without first implicitly determining whether or not the trial factor is indeed a factor \cite{Jones2008}, and so these implementations are not interesting from a computational point of view.  It is not yet clear to what extent these criticisms also apply to other techniques for implementing Gauss sums \cite{Wolk2009}.

\section{Other quantum information phenomena}
In this section I consider a range of phenomena and techniques from quantum information processing which have been demonstrated or explored using NMR techniques.  As was the case for quantum computation NMR is a useful technique for \textit{in-principle} demonstrations of many of these topics even though the lack of projective measurements and pure states means that the techniques cannot normally be used for anything genuinely useful.

\subsection{Quantum communication protocols}
Quantum communication protocols use quantum mechanical effects to transmit information between two or more parties in a way which is impossible with purely classical devices.  Perhaps the most important example is \textit{quantum cryptography} \cite{Gisin2002}, which permits two parties, usually called Alice and Bob, to securely establish a random key which can be used to encrypt a subsequent message, but many other examples are known.  NMR is peculiarly ill-suited to such applications, not only due to the lack of projective measurements and pure states, but also because it is effectively impossible to spatially separate individual qubits which are encoded as nuclear spins in a single molecule, but this has not prevented simple demonstration experiments from being performed.

To begin with, note that a qubit only has two distinguishable states, \ket{0} and \ket{1}, and so it might seem that only a single bit of classical information could be reliably transmitted by a single qubit.  (While superposition states appear to store more information, in that it takes more than a single bit of information to accurately describe them, this information cannot be accessed by measuring the state  \cite{Chefles2000}.)  \textit{Quantum dense coding} \cite{Bennett1992a} uses the peculiar properties of entangled states to enable two bits of classical information to be transmitted by a single quantum bit, which can normally only reliably transmit a single bit of information.

The protocol assumes that Alice and Bob each possess one of a pair of qubits in the entangled Bell state \ket{\psi^-}, and that Alice wishes to transmit a two bit message to Bob.  She can do this by performing one of four transformations, \textbf{1}, X, Y, or Z, to her qubit; as the two qubits are entangled this operation affects the whole two qubit state, converting it to \ket{\psi^-}, \ket{\phi^-}, \ket{\phi^+}, or \ket{\psi^+} respectively.  If Alice then sends her qubit to Bob he can determine which of these four Bell states the two qubits are in, and thus which message Alice wished to send.  This protocol has been implemented with a two qubit system \cite{Fang2000}, and a generalisation has been demonstrated in a three qubit system \cite{Wei2004}.

\textit{Quantum teleportation} \cite{Bennett1993} also uses entangled states, this time to transmit an unknown quantum state.  Suppose Alice has a qubit in some unknown state, which she wishes to send to Bob.  As she does not know the state she cannot simply send him a description, and it is furthermore impossible for her to obtain these details as it is impossible to accurately measure an unknown quantum state \cite{Chefles2000}.  Thus it might seem that the only way in which she can transfer her state to Bob is to physically transfer the qubit carrying the state.

Quantum teleportation, however, enables her to achieve this without physically transferring the qubit as long as she shares an entangled state with Bob and is able to send him two bits of classical information; her original copy is destroyed in the process.  This protocol was demonstrated in a three-spin system \cite{Nielsen1998}, and was indeed one of the earliest quantum protocols to be demonstrated by NMR.  This experiment relied on rapid $\textrm{T}_2$ decoherence of two \nuc{13}{C} nuclei to simulate the effects of quantum measurement, and emulated Bob's final state transformation step by post-processing, thus greatly simplifying the implementation.

\textit{Entanglement swapping} \cite{Bennett1992a,Zukowski1993,Pan1998} is a close relative of quantum teleportation, in which the quantum state to be teleported is itself entangled with another qubit.  As quantum teleportation transfers a perfect copy of the unknown quantum state, it will transfer this entanglement.  Thus if Alice shares an entangled state with Bob, and another such state with Charlie, she can teleport her entanglement such that Bob and Charlie now share an entangled state.  This approach can be generalised to distribute more complex forms of entanglement, sometimes called a quantum telephone exchange \cite{Bose1998}.  The scheme has been demonstrated in a four spin system \cite{Boulant2003}, using artificially imposed decoherence to implement the measurement step.  Entanglement transfer can also be achieved by replacing the teleportation stage with a pair of quantum \textsc{swap} gates \cite{Boulant2002}, but this is not normally considered a communication protocol as it requires coherent interactions between the three parties.

\textit{Remote state preparation} \cite{Pati2000,Bennett2001} is another generalisation of quantum teleportation, in which Alice wishes to transfer a known quantum state to Bob.  As she knows the state she could simply transmit an accurate description, but this requires transmitting two real numbers, such as the $\theta$ and $\phi$ coordinates of the state on the Bloch sphere.  She could instead teleport the state, requiring only two classical bits to be sent, as long as she shares an entangled state with Bob.  As she knows the state, however, the process can be slightly simplified, and the state can be transmitted using a single classical bit.  Furthermore it is not strictly necessary for Alice to even prepare a qubit in the known state, and the whole process can be performed using only the two qubits making up the entangled state; this simplified process has been demonstrated in a two qubit system \cite{Peng2003}.

\textit{Quantum random walks} \cite{Aharonov1993,Kempe2003} are quite different from their classical counterparts, as a result of quantum interference effects.  Both continuous time \cite{Du2003a} and discrete time \cite{Ryan2005} walks have been investigated by NMR.

\textit{Quantum chaos} occurs when a quantum system evolves under a Hamiltonian in such a way that its final state is hypersensitive to small perturbations in the dynamics \cite{Schack1993}.  A simple example, the quantum Baker's map \cite{Schack1993}, has been studied using a three-spin system \cite{Weinstein2002}.

Many quantum communication protocols can be recast using the idea of \textit{quantum games} \cite{Meyer1999,Eisert1999,Benjamin2001a,Patel2007}, with the participants in the protocol, and sometimes nature itself, taking the role of players.  There have been two explicit demonstrations of a traditional quantum game, the prisoner's dilemma, using two-spin \cite{Du2002} and three-spin \cite{Mitra2007} systems.

\subsection{Quantum Cloning}
It is a well known result in quantum information theory that a qubit in an unknown state cannot be accurately copied (\textit{cloned}) \cite{Wooters1982}.  It is, however, possible to prepare approximate copies \cite{Buzek1996}, and specific quantum networks for doing this have been described.  For example an optimal one-to-two cloning network \cite{Buzek1997} converts a single copy of a unknown quantum state \ket{\psi} into two approximate copies of the form
\begin{equation}
\mbox{$\frac{5}{6}$}\ket{\psi}\bra{\psi}+\mbox{$\frac{1}{6}$}\ket{\psi^\perp}\bra{\psi^\perp}=\mbox{$\frac{2}{3}$}\ket{\psi}\bra{\psi}+\mbox{$\frac{1}{3}$}\mathbf{1}/2
\end{equation}
where \ket{\psi^\perp} is the state orthogonal to \ket{\psi}. This description can be naturally generalised to mixed states, and such mixed state quantum cloning has been demonstrated by NMR \cite{Cummins2002}.  This implementation used a three-spin system, in which the initial spin is cloned onto two ancilla spins. The initial spin is left in a mixed state corresponding to an optimal implementation of a transpose operation on the original state, which is closely related to a universal-\NOT\ gate \cite{Buzek1999,Buzek2000}.

There are several other processes closely related to quantum cloning, such as \textit{cloning with prior partial information} \cite{Du2005}, which permits more accurate copies to be made if the initial state is limited to a subset of states, such as states on the Bloch sphere, Eq.~\ref{eq:qubitbs}, with a fixed co-latitude but unknown azimuth angle.  There has also been an implementation of \textit{asymmetric quantum cloning} \cite{Chen2007a} in a two-spin system; this process uses no ancilla spins, and allows the fidelity of the cloning to be arbitrarily divided between the two copies.

\subsection{Decoherence, error correction, and fault tolerance}\label{sec:ECDFS}
Decoherence processes, that is unwanted and uncontrolled interactions between individual qubits, or between qubits and the environment, will lead to errors in the values represented by these quantum states.  Although decoherence effects can be minimised by careful control of such interactions, such errors cannot be entirely avoided, and must be either suppressed or corrected \cite{Preskill1999}.  The two principal methods for doing this are the use of \textit{decoherence free subspaces} \cite{Zanardi1997,Duan1997,Lidar1998,Lidar2001,Lidar2001a} and \textit{quantum error correction} \cite{Shor1995,Steane1996,Calderbank1996,Laflamme1996}.  These can be thought of as quantum communication protocols, allowing a quantum state to be accurately transmitted from the present to some later time.  A further generalisation of this is \textit{fault tolerant computation} \cite{Knill1998b,Gottesman1998,Bacon2000}, which considers the effects of decoherence during the application of logic gates, particularly those logic gates used to implement decoherence free subspaces and error correction.  All these methods rely on replacing a single physical qubit with a \textit{logical qubit}, encoded with the help of multiple ancilla qubits.

When considering methods for handling decoherence it is useful to distinguish between two major types: phase decoherence, which can be considered as the random application of Z gates to the state, and spin flip decoherence, which arises from random X gates; more complex decoherence processes can be composed as a combination of these fundamental operations \cite{NCbook}, and so it normally suffices to protect a system against these two types of decoherence.  (Note that phase decoherence is essentially equivalent to $\textrm{T}_2$ relaxation, while spin flip decoherence is not quite the same as $\textrm{T}_1$ relaxation.)  The two processes are closely related, as a random X gate is equivalent to a random Z gate surrounded by Hadamard gates, and so a method which protects against phase decoherence can easily be converted to one that protects against spin flips, and \textit{vice versa}.  Protection against both sorts of decoherence simultaneously can be achieved either directly or by nesting methods that protect against the individual errors; the latter approach is often simpler to implement but requires a larger overhead in ancilla qubits.

Decoherence free subspaces use correlations between the decoherence of individual qubits to design multi-qubit states which are (ideally) unaffected by decoherence.  Although it is common to assume that spins relax independently, this is rarely the case as discussed in Section~\ref{sec:DLS} above.  These effects can be considered as implicit demonstrations of decoherence free subspaces, a clear example being the use of zero-quantum coherences to observe NMR signals in the presence of intense local magnetic field inhomogeneities \cite{Dixon1994}.  These simple schemes usually only involve protection of the system against transverse relaxation. Using the theory of decoherence free subspaces \cite{Zanardi1997,Duan1997,Lidar1998,Lidar2001,Lidar2001a} it is possible to design systems which are simultaneously protected against correlated relaxation of all kinds by using multiple spins to encode a single qubit.

An explicit demonstration has been performed \cite{Viola2001} in which a single qubit encoded on three \nuc{13}{C} nuclei was shown to be resistant against artificially imposed decoherence (see Section~\ref{sec:forceddecoherence}), and this system was subsequently studied in detail \cite{Fortunato2003}. A simpler system in which a qubit is encoded in two spins, providing resistance only to transverse relaxation, has also been studied \cite{Fortunato2002a}, and the use of identical spins in $\textrm{CH}_n$ systems has been explored \cite{Wei2005}.  However, decoherence free subspaces are only really useful when multiple qubits are simultaneously protected.  To date a four-spin system has been used to protect two qubits against artificially applied spin flip decoherence \cite{Ollerenshaw2003}, and it was shown that the resulting device could perform a quantum algorithm, while a similar system not using decoherence free subspaces could not.  The behaviour of systems of this kind has been studied in considerable detail \cite{Hodges2007,Cappellaro2006}.

While correlated errors can be suppressed using decoherence free subspaces, uncorrelated (independent) errors must be detected and corrected.  In classical information processing this problem is well understood \cite{FCbook}, and the solution relies on redundancy, keeping multiple copies of each bit so that disagreements between copies can be identified and reconciled.  The simplest approach is to keep three copies of every bit; should these three copies ever disagree then the single disagreeing bit can be reset to the majority value.  This approach assumes that the probability of two bits being wrong is negligible, which will be reasonable if the errors occur independently and infrequently, so that the probability of two errors is much smaller than the probability of one.  If the error rate is too high for this approximation to be used then more sophisticated methods, using larger numbers of copies, can be used.  Similarly, if the error rate is very low then it is possible to use parity checks to correct errors with a smaller number of ancilla bits.

At first sight it is difficult to see how these ideas could be used in quantum computation, as they appear to rely on measuring the bits to detect differences between copies, and this cannot be done for bits in superposition states without causing the superpositions to collapse.  It is, however, not actually necessary to measure the bits to detect errors: it suffices to detect whether two bits are the same or different, without actually determining their states  \cite{Shor1995,Steane1996,Calderbank1996,Laflamme1996}, and the erroneous copy can then be rotated back into place.

Quantum error correction has been implemented using a three-spin system \cite{Cory1998} to protect against phase decoherence.  A related demonstration used a two-spin system to implement detection (but not correction) of phase errors \cite{Leung1999}, permitting a detailed characterisation of the process; a detailed analysis of the three-spin system was subsequently performed \cite{Sharf2000a}.   Protecting a single qubit against arbitrary decoherence requires four ancilla qubits \cite{Laflamme1996}, and this has been demonstrated using a five-spin system \cite{Knill2001}.  Quantum error correction has also been successfully combined with decoherence free subspaces using a four-spin system \cite{Boulant2005}.

All these demonstrations have only implemented a \textit{single round} of error correction.  In order to preserve a qubit state indefinitely it is necessary that the qubit state be corrected repeatedly, and this in turn requires that the ancilla qubits used to detect errors be reset back to their initial state after each round.  As previously noted, the ability to reset qubits during a computation is not available in NMR implementations, and so full error correction is currently impossible.  Error correction has also been implemented in ion traps \cite{Chiaverini2004,Steane2004}, where resetting qubits is theoretically possible; however as yet no complete demonstration of multiple round error correction has been performed.

\subsection{The quantum Zeno effect}
The \textit{quantum Zeno effect} \cite{Misra1977,Itano1990,Nakazato1996,Home1997} refers to the fact it is possible to suppress the coherent evolution of a quantum system by making frequent measurements, which project the quantum system onto its eigenstates, during the evolution process.  As an example consider a \NOT\ gate applied to a qubit starting in state \ket{0}.  (In NMR terms this corresponds to a $180^\circ_x$ rotation applied to an isolated spin starting in the state $I_z$.)  During the evolution the state is described by
\begin{equation}
\ket{\psi(t)}=\cos(\omega{t}/2)\ket{0}+i\sin(\omega{t}/2)\ket{1}
\end{equation}
with nutation rate $\omega=\pi/t_{180}$.  Suppose, however, that measurements are made in the $\{\ket{0},\,\ket{1}\}$ basis at time intervals $\tau$: each measurement will project the system onto either state $\ket{0}$, with probability $\cos^2(\omega\tau/2)$, or state $\ket{1}$, with probability $\sin^2(\omega\tau/2)$.  If measurements are made rapidly, such that $\omega\tau\ll1$, the system will almost always be found after the first measurement in the initial state $\ket{0}$.  Subsequent evolution and measurements will have the same effect, with the system being returned to the initial state each time, so frequent measurements can effectively suppress the nutation.

It is not immediately obvious how NMR can be used to demonstrate a quantum Zeno effect, as the standard NMR measurement process is not a true projective measurement.  It is, however, possible to simulate the effects of projective measurements by forced decoherence as described in Section~\ref{sec:forceddecoherence} above.  This approach has been used to demonstrate the quantum Zeno effect using a single \nuc{1}{H} spin \cite{Xiao2006a}.  The effects of measurement strength were also explored in a two-spin system \cite{Xiao2006a}.  An analogy can be drawn between the Zeno effect and the behaviour of frequency selective pulses \cite{Xiao2006a}, and more generally between Zeno effects and dynamical decoupling sequences \cite{Facchi2004}.

\subsection{Multi-spin entanglement}
Although high temperature pseudo-pure spins states cannot be truly entangled, there has been interest in preparing pseudo-entangled states \cite{Jones2001a}, that is pseudo-pure states whose pure part is entangled.  This is sometimes seen as a test of quantum computing \cite{Knill2000}, as a general purpose quantum computer should, among other things, be able to produce highly entangled states in an $n$-qubit system.

In a two qubit system the maximally entangled states are the Bell states, but in larger systems the situation is less simple, as different entanglement measures can lead to different conclusions.  With three qubits the states considered most frequently are the Greenberger--Horne--Zeilinger (GHZ) states \cite{Greenberger1989,Mermin1990}
\begin{equation}
\ket{\psi_{\textrm{GHZ}}}=(\ket{000}\pm\ket{111})/\sqrt{2}
\end{equation}
and the W state \cite{Dur2000}
\begin{equation}
\ket{\psi_{W}}=(\ket{001}+\ket{010}+\ket{100})/\sqrt{3}
\end{equation}
although a more complete classification of states is available \cite{Acin2001}.  GHZ states were produced in early experiments \cite{Laflamme1998}, although in this case a pseudo-pure state was not initially prepared, with the presence of the GHZ component being detected by state tomography.  W states have been produced for studying disentanglement erasers \cite{Teklemariam2001,Teklemariam2002,Teklemariam2003}.

Studies on larger spin system have mostly concentrated on cat states, which are analogous to GHZ states; notable results include 7-spin \cite{Knill2000} and 10-spin \cite{Negrevergne2006} states, as well as partial control of a 12-spin cat state \cite{Negrevergne2006}.  Cat states have also been explored in the context of entanglement assisted magnetic field sensing, where 10-spin \cite{Jones2009} and 13-spin \cite{Simmons2010} states have been prepared with global control.  The similarities between cat states and maximal quantum coherence have been explored in liquid crystal \cite{Lee2004,Lee2005,Lee2005a,Lee2007} and solid state \cite{Warren1980,Doronin2003,Doronin2007} systems.

\section{Conclusions}
Since the first tentative suggestions that NMR could be used to implement quantum computation \cite{Cory1996} there has been quite extraordinary progress.  In this final section I consider briefly how the field has stood up to early predictions, what has been achieved on the way, and where the future might lie.

Since the very beginning of the field it has been understood that NMR cannot provide a practical technology for implementing genuinely useful quantum computations, partly because of the exponentially small signal size available as the size of the spin system is increased \cite{Warren1997}, but also because of the sheer difficulty of implementing quantum circuits in larger molecules \cite{Jones1999b}.  The general belief was that systems with less than five qubits would be straightforward, but that larger systems would become increasingly difficult, with a limit between ten and twenty (just possibly thirty) qubits, and this has proved broadly correct, with the largest spin-systems explored to date containing fourteen nuclei \cite{Negrevergne2006}.

As expected \cite{Jones1999b}, NMR remained the leading technology for implementing quantum algorithms for around five to ten years.  This reflects the extreme experimental challenges involved in implementing quantum computations with, for example, trapped ions.  NMR remains by far the simplest technology for building small demonstration devices, and is so far the only technology used to implement Shor's quantum factoring algorithm.  However the insurmountable difficulties in building genuinely useful computers with liquid state NMR have led some authors to question whether there is any point in pursuing the field.  If we cannot possibly reach our desired destination, what is the point in setting out?  This question might seem reasonable, but its premise is too restrictive: sometimes one walks not to reach a destination, but to observe the scenery along the way, and the pursuit of NMR quantum computation has thrown up some surprising sights.

Beyond the sheer interest of these ideas, one must also consider the potential role of NMR quantum computing in catalysing \textit{technology transfer} between conventional NMR and quantum information processing.  To date NMR has largely been the donor of ideas, with the use of composite pulses to suppress systematic errors being a particularly important example.  Large scale quantum computers require single qubit rotations to be carried out with extremely high precision, and without the use of composite pulses it is hard to see how these could in fact be implemented experimentally.  The quantum computing world has extended these ideas with the concept of arbitrarily accurate composite pulses \cite{Brown2004}, but it seems unlikely that these will find application in conventional NMR where ultra-high precision rotations are unnecessary.  However the near ubiquity of techniques to suppress pulse-length errors in NMR quantum computation should provide inspiration to more conventional work: the BB1 composite pulse \cite{Wimperis1994} is such a good replacement for a simple hard pulse that there is no reason why it should not be used completely routinely in most simple pulse-acquire experiments.

Dynamical decoupling \cite{Viola1999} is another area where the two fields may have much to offer one another, but it is perhaps not sensible to make too many firm and detailed predictions.  Implementing quantum computing has enriched NMR by setting a series of challenging technical demands, which NMR has risen to magnificently.  In return NMR has enriched quantum computation by proving that implementing complex unitary transformations can actually be quite straightforward, and by raising uncomfortable questions about the role of entanglement in mixed states.  Most of the ``easy'' experiments have probably been performed by now, but there is still plenty to do, and the transfer of many of these experiments from liquid state NMR to more demanding fields, such as solid state ENDOR where high purity states can be reached directly by cooling, is likely to keep us busy for some time to come.

%\section*{Acknowledgements}
\ack
I thank the UK EPSRC and BBSRC for financial support and Sabieh Anwar, Arzhang Ardavan, Jonathan Baugh, Ruth Dixon, Joe Fitzsimons, Dieter Jaksch, Minaru Kawamura,
Michael Mehring, John Morton, Ben Rowland, Thomas Schulte-Herbr\"uggen, Stephanie Simmons, Andrew Steane, Chris Timpson, Vlatko Vedral and Li Xiao for helpful discussions.

\bibliographystyle{elsart-num-jaj}
\bibliography{all}

\end{document}